\documentclass[12pt,draftclsnofoot,onecolumn]{IEEEtran}
\usepackage{amssymb,amsthm,amsmath,amsfonts} 
\usepackage{mathtools} 
\usepackage{dsfont} 
\usepackage[binary-units]{siunitx} 
\usepackage{graphicx} 
\usepackage{caption} 
\usepackage{subcaption} 
\usepackage{mathrsfs,bm,stmaryrd} 
\usepackage[final]{pdfpages} 
\usepackage{tikz} 
\usetikzlibrary{tikzmark} 
\usetikzlibrary{patterns,positioning}
\usepackage{pgfplots} 
\usepackage{cite} 
\usepackage{color} 

\newcommand{\mat}[1]{\mathbf{#1}} 
\DeclareMathOperator{\Tr}{Tr} 

\newtheorem{proposition}{Proposition}
\newtheorem{corollary}{Corollary}

\begin{document}

This work has been submitted to the IEEE for possible publication.  Copyright may be transferred without notice, after which this version may no longer be accessible.
\newpage

\title{An Information-Theoretic Analysis of the Gaussian Multicast Channel with Interactive User Cooperation}
\author{\IEEEauthorblockN{Victor~Exposito\IEEEauthorrefmark{1}\IEEEauthorrefmark{2},\IEEEmembership{~Student Member,~IEEE}, Sheng~Yang\IEEEauthorrefmark{1},\IEEEmembership{~Member,~IEEE}, Nicolas~Gresset\IEEEauthorrefmark{2},\IEEEmembership{~Member,~IEEE}}
\thanks{\IEEEauthorblockA{\IEEEauthorrefmark{1}Laboratoire des Signaux et Syst\`emes~(L2S, UMR CNRS 8506), CentraleSup\'elec, Universit\'e Paris-Sud, Universit\'e Paris-Saclay, Gif-sur-Yvette, France.}}
\thanks{\IEEEauthorblockA{\IEEEauthorrefmark{2}Mitsubishi Electric R\&D Centre Europe, Rennes, France.}}
\thanks{Email: \{victor.exposito,sheng.yang\}@centralesupelec.fr, \{v.exposito,n.gresset\}@fr.merce.mee.com}}
\maketitle
\IEEEpeerreviewmaketitle

\begin{abstract}
	We consider the transmission of a common message from a transmitter to three receivers over a broadcast channel, referred to as a multicast channel in this case. All the receivers are allowed to cooperate with each other over full-duplex bi-directional non-orthogonal cooperation links. We investigate the information-theoretic upper and lower bounds on the transmission rate. In particular, we propose a three-receiver fully interactive cooperation scheme~(3FC) based on superpositions of CF and DF at the receivers. In the 3FC scheme, the receivers interactively perform compress-forward~(CF) simultaneously to initiate the scheme, and then decode-forward~(DF) sequentially to allow a correlation of each layer of the DF superposition in cooperation with the transmitter toward the next receiver in the chain to improve the achievable rate. The analysis leads to a closed-form expression that allows for numerical evaluation, and also gives some insight on key points to design interactive schemes. The numerical results provided in the Gaussian case show that the proposed scheme outperforms existing schemes and show the benefit of interaction.
\end{abstract}


\section{Introduction}
\label{sec:Introduction}

In nowadays and future wireless communication systems, an intensification of the request of content delivery in increasingly denser and more heterogeneous networks is taking place. This escalation leads, among many, to a spectrum crunch or an interference intensification. To tackle one part of this problem, we focus on the multicast channel~(MC) in which one transmitter broadcasts a common message intended to a whole group of users. The MC models a wide range of scenarios, such as the streaming of multimedia content, the spreading of data in public safety or industrial networks, and the control signaling in sensor networks~\cite{article41}. To ensure that the transmission rate is not limited by the weakest user in terms of channel quality, different solutions have been proposed using multilayer strategies or massive multiple-input multiple-output~(MIMO)~\cite{unpublished5}. However, if all users wish to obtain the same content quality, the weakest user would set the rate and/or require a disproportionate amount of resource, and thus impact the whole group. With the recent study of device-to-device~(D2D) mechanisms in standards~\cite{misc2,misc3}, user cooperation in close proximity becomes possible and would benefit to all users by ensuring the same content quality while maintaining a low cost in terms of amount of resource and energy~\cite{inproceedings3}.

In this work, we investigate the broadcast channel~(BC) with one transmitter sending a common message to three receivers, also called a MC in this case. The receivers can cooperate through a cooperation link. The goal is to characterize the benefits of cooperation in terms of achievable rate through an information-theoretic analysis. Since the receivers also transmit signals through the cooperation link, our channel is a mixture of the MC and the relay channel~(RC). The choice of three receivers comes from the fact that it is the smallest size to clearly show the core idea of our scheme compared to others while remaining tractable for the information-theoretic analysis as well as the numerical results.

The RCs~\cite{article26} have been well studied in the past. In~\cite{article30}, two relaying strategies called compress-forward~(CF) and decode-forward~(DF) were proposed for the basic three-node network. These schemes were then extended to larger networks~\cite{article23,inproceedings13,article32,article33,article31,unpublished10}. Among them, a particularly interesting scheme is the noisy network coding~(NNC)~\cite{inproceedings13,article32} and its more recent variant called the short message NNC~(SNNC)\footnote{Since the achievable rate of the SNNC (with backward decoding or sliding window decoding) is equal to the one of the NNC (with joint decoding), we do not distinguish between the NNC and SNNC schemes hereafter.}~\cite{article33}. The NNC readily applies for multicast networks and achieves within a constant gap to the capacity. Refinements called the SNNC with a DF option~(SNNC-DF)~\cite{article33} and the SNNC with rate-splitting~\cite{unpublished10} were also developed for unicast. The capacity of the BC with cooperation, even in the case of two receivers, remains unknown in general, except for special cases such as the physically degraded main channel~\cite{article14}. The setup for two receivers has been partially studied in~\cite{article14}, referred to as BCs with cooperative decoders, and in~\cite{article19,article21}, referred to as relay BCs. A BC with orthogonal cooperation links was considered in~\cite{article14}. In~\cite{article19,article21}, although the cooperation links are not restricted to be orthogonal, the authors assumed that either the main channel is degraded or the cooperation link is uni-directional. It is worth noting that achievable rate regions of both common and private messages were provided in~\cite{article14,article19,article21}. In our previous work~\cite{ownpublications1} we generalized the results of~\cite{article14,article19,article21} by studying the full-duplex bi-directional non-orthogonal cooperation link counterpart for the MCs. In that work, we proposed the two-round interactive receiver cooperation scheme~(2RC), in which one receiver uses CF toward the other one which in turns uses DF back to the first one. It turned out that the 2RC outperforms both the NNC and DF cooperation, which shows the benefit of interaction between compression and decoding.

To investigate the benefit of such an interaction in a larger network, we propose a new three-receiver fully interactive cooperation scheme~(3FC). In the proposed scheme, the transmitter multicasts a short message and then sequentially performs the following three steps, 1)~receivers~2 and~3 use CF toward receiver~1 in the first block, 2)~receiver~1 cooperates with the transmitter by using DF toward receiver~2 in the second block, and 3)~receivers~1 and~2 cooperate with the transmitter by using DF toward receiver~3 in the third block. At this point, the scheme ends for this message, giving a latency of 3~blocks if sliding window decoding is implemented. This sequence is repeated identically in each block until the end of the scheme. The same holds for receivers~1, 2, and~3 exchanging roles. We present the cutset upper bound and three lower bounds for the MC with receiver cooperation, two of which are derived from existing results in the literature (``no cooperation" and NNC schemes), and the third one is a special case of our proposed scheme, which we call the three-receiver partially interactive cooperation scheme~(3PC). Note that the 2RC is a special case of the 3PC, and that the 3PC is a special case of the proposed 3FC. The 3PC is presented to show the importance of full cooperation, i.e., each receiver should contribute to achieve a better performance.

We believe that the results presented and discussed in this paper are new for several reasons. First, the 3FC scheme exploits the interaction between different nodes in terms of the order of compression and decoding according to the channel condition, which has not been studied before to the best of the authors' knowledge. Numerical results show that superior rate performances can be obtained as compared to the state of the art. Second, the bounds presented in this work are general and can be applied to different channel configurations including the orthogonal and half-duplex cases. As such, we do not need to explicitly construct different schemes for the aforementioned settings, as is usually done in many existing works. Third, the structure of the 3FC scheme gives some new insights that we believe are key to design practical interactive schemes. Moreover, a suboptimal equivalent could be further studied by using existing tools of the literature, as discussed throughout the paper.

In the literature, many protocols and practical schemes~\cite{inproceedings14,article35} have been proposed in the D2D area. Some schemes are developed to opportunistically use D2D links~\cite{unpublished6,article20}, and do not require all users to decode the message. Asymptotic behavior of large-scale random wireless networks have been studied using stochastic geometry~\cite{unpublished7,unpublished8,article37}, however, the transmitters perform a simple repetition protocol for the purpose of tractability, which is in general suboptimal and leads to a low spectral efficiency. Other metrics than the achievable rate can be considered, such as the network lifetime~\cite{article38} under which the network has to be working for the longest possible time.

The remainder of the paper is organized as follows. Sec.~\ref{sec:System Model} introduces the system model and the Single-Input Single-Output~(SISO) Gaussian MC as a special case. In Sec.~\ref{sec:Proposed Scheme} we present the 3FC, derive its special cases 3PC and 2RC. Numerical results for the SISO Gaussian MC are provided in Sec.~\ref{sec:Numerical Results} to show that the 3FC scheme surpasses existing schemes regarding the achievable rate and is thus a good generalization of the results of~\cite{ownpublications1}. We also further explain the terms of the bound, and underline the importance of the structure of the 3FC. Finally, we conclude the paper and discuss the generalized structure of our scheme in Sec.~\ref{sec:Summary and Discussion}.

We use the following notations throughout the paper. We denote random variables with upper case letters and their realizations with the corresponding lower case letters. The signals sent and received by the receiver~$k$ are denoted respectively by $X_k$ and $Y_k$, the compressed version of $Y_k$ is $\tilde{Y}_k$, and the decoded version of $Y_k$ is $\hat{Y}_k$. The mutual information between $X$ and $Y$ given $Z$ is denoted by $I(X;Y\vert Z)$. The $n$-sequence $1\leq n$ sent by receiver~$k$ is denoted $x_k^n$. The discrete interval $[i:j]=\left\{i,i+1,\ldots,j\right\}$ is defined for a pair of integers $i\leq j$. The notation $k\neq l\neq q$ means $k\in[1:3],\ l\in[1:3]\setminus\{k\},\ q\in[1:3]\setminus\{k,l\}$. The logarithms $\log(\cdot)$ are to base~2 and $\mathscr{C}(x)=\log(1+x)$.

\section{System Model}
\label{sec:System Model}

\begin{figure}[t]
	\centering
	\includegraphics[width=0.4\columnwidth]{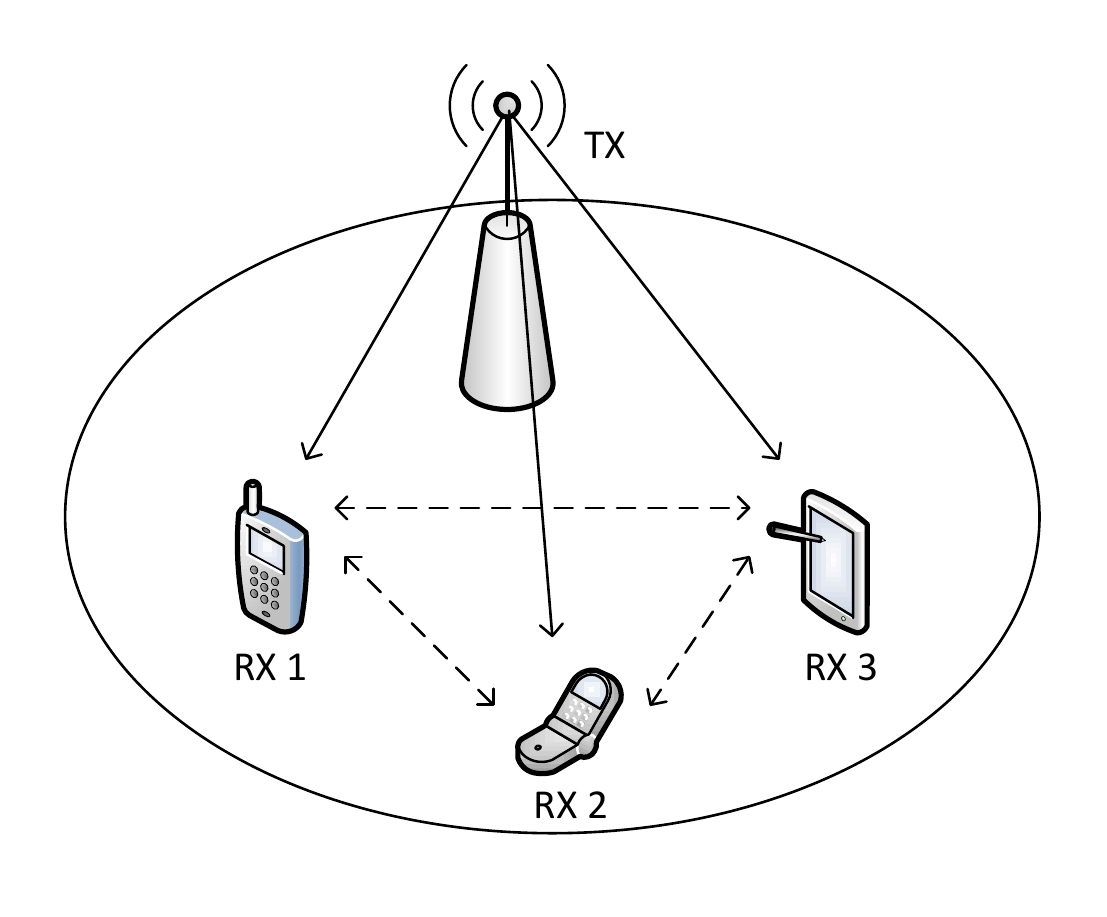}
	\caption{MC system with receiver cooperation.}
	\label{fig:MC system with receiver cooperation}
\end{figure}

We consider a MC with receiver cooperation, where one transmitter sends the same information to three receivers through the main channel as represented in Fig.~\ref{fig:MC system with receiver cooperation}. As a special case of this model, the SISO Gaussian channel~(Gaussian inputs and noises) at an instant~$i$ is described by
\begin{align}
  y_{ki}=h_kx_i+h_{lk}x_{li}+h_{qk}x_{qi}+z_{ki},\ \forall i\in[1:n]\label{eq:AWGN k}
\end{align}
for the receiver indices $k\neq l\neq q$; $x$ is the source signal, $x_k$ is the signal transmitted by receiver~$k$, and $y_k$ is the received signal at receiver~$k$; $h_l,\ h_{kl}\in\mathbb{C}$ are the channel coefficients from the source and from receiver~$k$ to~$l$, respectively; $z_{k}\sim\mathcal{CN}(0,\sigma^2)$ is the additive white Gaussian noise~(AWGN) at receiver~$k$, which is assumed to be independent across resources and receivers. We assume that the channel coefficients are constant and known globally at every node, which corresponds to the low-mobility scenario where the state information can be disseminated reliably. For simplicity, the same average power constraint is imposed for every emitting nodes, i.e., $\sum_{i=1}^{n}\vert x_i\vert^2\leq nP,\ \sum_{i=1}^{n}\vert x_{ki}\vert^2\leq nP,\ k\in[1:3]$. As such, the signal-to-noise ratios~(SNR) of the main channels are $\mathsf{SNR}_k=\vert h_k\vert^2\frac{P}{\sigma^2},\ k\in[1:3]$, and those of the cooperative links are $\mathsf{SNR}_{kl}=\vert h_{kl}\vert^2\frac{P}{\sigma^2},\ k\in[1:3],\ l\in[1:3]\setminus\{k\}$.

Instead of investigating this channel directly, we consider the more general class of non-orthogonal stationary memoryless channels. In this general model, the three receivers can cooperate with each other in full-duplex, i.e., they can transmit and receive simultaneously, through bi-directional non-orthogonal cooperation links. This setup includes, 1)~the cooperation links orthogonal to the main channel, orthogonal links being either physically separated medium, e.g., using different transmission technologies over different resources, or created with artificial orthogonalization, e.g., in time or frequency, and 2)~the half-duplex mode if the receivers transmit and receive at a different time. The current channel belongs to a class of stationary memoryless channels $(\mathcal{X}\times\mathcal{X}_1\times\mathcal{X}_2\times\mathcal{X}_3,p(y_1,y_2,y_3\vert x,x_1,x_2,x_3),\mathcal{Y}_1\times\mathcal{Y}_2\times\mathcal{Y}_3)$, defined as $p(y_1^n,y_2^n,y_3^n\vert x^n,x_1^n,x_2^n,x_3^n)=\prod_{i=1}^np(y_{1i},y_{2i},y_{3i}\vert x_i,x_{1i},x_{2i},x_{3i})$ where $x^n\in\mathcal{X}^n,\ x_k^n\in\mathcal{X}_k^n,\ k\in[1:3]$, and $y_k^n\in\mathcal{Y}_k^n$. The probability distribution of the channel is known at every node (perfect channel state information~(CSI) at the transmitter and receivers) by assumption. The common message $M$ is assumed to be uniformly distributed in $\mathcal{M}\triangleq[1:2^{nR}]$ where $R$ is the number of bits per channel use. An encoder at the transmitter side is a map $f_i^{(n)}$ from the message $M$ to the sequence of input symbols $x^n$, an encoder at the receiver~$k$, is a sequence of maps $\{f_{ki}^{(n)}\}_i$ from the past received symbols $y_k^{i-1}$ to the transmitted symbol $x_{ki}$. A decoder at the receiver~$k$ is a map $\{g_{ki}^{(n)}\}_i$ from the received sequence $y_k^n$ to $\hat{M}_k\in\mathcal{M}$. The probability of error is defined as $P_e^{(n)}\triangleq\Pr(\cup_{k=1}^3\hat{M}_k\neq M)$. Finally, a rate $R$ is achievable if there exist a sequence of encoders/decoders $\left(f_i^{(n)},\{f_{1i}^{(n)}\}_i,\{f_{2i}^{(n)}\}_i,\{f_{3i}^{(n)}\}_i,\{g_{1i}^{(n)}\}_i,\{g_{2i}^{(n)}\}_i,\{g_{3i}^{(n)}\}_i\right)$ such that $P_e^{(n)}\rightarrow0$ as $n\rightarrow\infty$. Note that we obtain an orthogonal channel if, 1)~we split $\mathcal{Y}_k=\mathcal{Y}_k^{\textrm{m}}\times\mathcal{Y}_k^{\textrm{c}}$ between the main channel and the cooperation links, 2)~we split $Y_k=(Y_k^{\textrm{m}},Y_k^{\textrm{c}})$ with $Y_k^{\textrm{m}}\in\mathcal{Y}_k^{\textrm{m}},\ Y_k^{\textrm{c}}\in\mathcal{Y}_k^{\textrm{c}},\ k\in[1:3]$, and 3)~we have
\begin{align}
	p(y_1,y_2,y_3\vert x,x_1,x_2,x_3)=p(y_1^{\textrm{m}},y_2^{\textrm{m}},y_3^{\textrm{m}}\vert x)\prod_{k\in[1:3]}p(y_k^{\textrm{c}}\vert x_{[1:3]\setminus\{k\}}),
\end{align}
i.e., the received signals from the main channel are independent of the received signals from the cooperation links. The information-theoretic bounds derived under those general classes of channels can be specialized for any stationary memoryless channels compliant to the corresponding requirements. Contrary to~\cite{ownpublications1}, in which the special cases were the orthogonal case (noiseless cooperation links of finite capacity), the SISO Gaussian case, and the MISO Gaussian case, in the present paper, the orthogonal case is not studied since it is straightforward, and the MISO Gaussian case is not studied neither, due to its high number of parameters to optimize, and since the conclusions would certainly be very similar to the ones already presented.

Note that in~\eqref{eq:AWGN k}, self-interference is not considered at the receivers, i.e., it can be removed using the perfect CSI assumption. In practice, self-interference could be dealt with by data processing or resource decoupling. A mix of antenna separation and of analog/digital cancellation is studied in~\cite{inproceedings16,article43}, and it is shown experimentally that those techniques can suppress from~\SIrange{40}{80}{\decibel} of self-interference using only off-the-shelf technologies and that it is sufficient to support full-duplex wireless communication. Moreover, while it is shown in~\cite{inproceedings16,inproceedings15} that for an identical amount of resource, full-duplex (subject to self-interference) does not always outperform half-duplex (limited by the transmit/receive time allocation) from an achievable rate or degree-of-freedom~(DoF) point of view, we only consider full-duplex in our calculus since the result in half-duplex can be derived as a special case. In~\cite{ownpublications2}, we have shown that the 2RC full-duplex scheme always outperforms the 2RC half-duplex schemes since the 2RC scheme does not present any kind of interference due to its construction~\cite{ownpublications1}. The conclusions could be different in some cases for the 3FC scheme since there is more information flowing through the network and because the sliding window decoding does not permit to always remove all the superposition layers that are neither of interest nor already known for a given receiver. Another model to deal with full-duplex~\cite{unpublished11} in a multihop unicast relaying scheme uses virtual full-duplex relay channels. In this model the receive and transmit antennas of the relays belong to physically separated nodes. Thus, one relay is split into two nodes (by considering that self-interference can be dealt with) that can perform half-duplex relaying and that are used alternatively in transmit or receive modes. The present paper does not address higher protocol-level issues that may arise in practice. We concentrate on the information-theoretic bounds to design a good cooperation scheme, while ensuring that a suboptimal equivalent could be implemented, and with the anticipation that a higher level overhead will be negligible compared to the gains reported herein.

\section{Proposed Scheme}
\label{sec:Proposed Scheme}

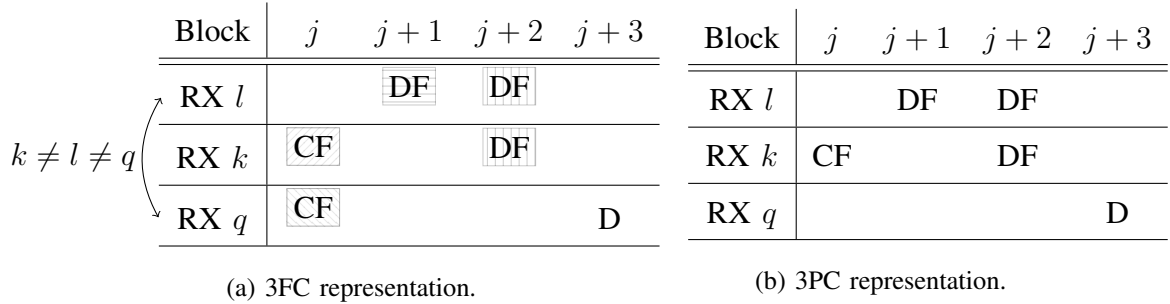
\begin{figure}[t]
	\centering
	\begin{subfigure}[t]{.31\textwidth}
		\centering
		\begin{tabular}{c|cccc}
			Block&$j$&$j+1$&$j+2$&$j+3$\\\hline\hline
			\tikzmark{j-l}RX~$l$&&\begin{tikzpicture}[stripeal/.style={draw, pattern=north east lines},stripear/.style={draw, pattern=north west lines},stripea/.style={draw, pattern=horizontal lines},stripel/.style={draw, pattern=vertical lines}]
				\coordinate(belowbelowleft)at(-0.35,-0.25);
				\coordinate(aboveright)at(0.35,0.25);
				\draw[stripea,opacity=0.2](belowbelowleft)rectangle(aboveright);
				\node(belowbelowleft){DF};
			\end{tikzpicture}&\begin{tikzpicture}[stripeal/.style={draw, pattern=north east lines},stripear/.style={draw, pattern=north west lines},stripea/.style={draw, pattern=horizontal lines},stripel/.style={draw, pattern=vertical lines}]
				\coordinate(belowbelowleft)at(-0.35,-0.25);
				\coordinate(aboveright)at(0.35,0.25);
				\draw[stripel,opacity=0.2](belowbelowleft)rectangle(aboveright);
				\node(belowbelowleft){DF};
			\end{tikzpicture}&\\\hline
			\tikzmark{j-k}RX~$k$&\begin{tikzpicture}[stripeal/.style={draw, pattern=north east lines},stripear/.style={draw, pattern=north west lines},stripea/.style={draw, pattern=horizontal lines},stripel/.style={draw, pattern=vertical lines}]
				\coordinate(belowbelowleft)at(-0.35,-0.25);
				\coordinate(aboveright)at(0.35,0.25);
				\draw[stripeal,opacity=0.3](belowbelowleft)rectangle(aboveright);
				\node(belowbelowleft){CF};
			\end{tikzpicture}&&\begin{tikzpicture}[stripeal/.style={draw, pattern=north east lines},stripear/.style={draw, pattern=north west lines},stripea/.style={draw, pattern=horizontal lines},stripel/.style={draw, pattern=vertical lines}]
				\coordinate(belowbelowleft)at(-0.35,-0.25);
				\coordinate(aboveright)at(0.35,0.25);
				\draw[stripel,opacity=0.2](belowbelowleft)rectangle(aboveright);
				\node(belowbelowleft){DF};
			\end{tikzpicture}&\\\hline
			\tikzmark{j-q}RX~$q$&\begin{tikzpicture}[stripeal/.style={draw, pattern=north east lines},stripear/.style={draw, pattern=north west lines},stripea/.style={draw, pattern=horizontal lines},stripel/.style={draw, pattern=vertical lines}]
				\coordinate(belowbelowleft)at(-0.35,-0.25);
				\coordinate(aboveright)at(0.35,0.25);
				\draw[stripear,opacity=0.3](belowbelowleft)rectangle(aboveright);
				\node(belowbelowleft){CF};
			\end{tikzpicture}&&&D\\\hline
		\end{tabular}
		\begin{tikzpicture}[overlay,remember picture,yshift=.25pt,shorten >=.5pt,shorten <=.5pt]
			\draw[<->]([xshift=-7pt,yshift=4pt]{pic cs:j-q})to[bend left]([xshift=-7pt,yshift=4pt]{pic cs:j-l});
			\draw([xshift=-11pt,yshift=4pt]{pic cs:j-k})node[left]{$k\neq l\neq q$};
		\end{tikzpicture}
		\caption{3FC representation.}
		\label{fig:The MC with receiver cooperation for the 3FC and 3PC schemes - 3FC representation}
  \end{subfigure}
	~
	~
	~
	~
	~
	~
	\begin{subfigure}[t]{.31\textwidth}
		\centering
		\begin{tabular}{c|cccc}
			Block&$j$&$j+1$&$j+2$&$j+3$\\\hline\hline
			RX~$l$&&DF&DF&\\\hline
			RX~$k$&CF&&DF&\\\hline
			RX~$q$&&&&D\\\hline
		\end{tabular}
		\caption{3PC representation.}
		\label{fig:The MC with receiver cooperation for the 3FC and 3PC schemes - 3PC representation}
  \end{subfigure}
	\caption{The MC with receiver cooperation for the 3FC and 3PC schemes. The oblique stripes represent CF operations while vertical and horizontal ones represent DF operations.}
	\label{fig:The MC with receiver cooperation for the 3FC and 3PC schemes}
\end{figure}

In this section, we first present the intuition behind the proposed 3FC scheme, illustrated in a simplified manner in Fig.~\ref{fig:The MC with receiver cooperation for the 3FC and 3PC schemes - 3FC representation}, and present the corresponding bounds. The coding scheme and techniques used to prove Prop.~\ref{prop:3FC} are provided in Appx~\ref{appx:Description of the Proposed Scheme}. We then derive as a special case the 3PC scheme, illustrated in Fig.~\ref{fig:The MC with receiver cooperation for the 3FC and 3PC schemes - 3PC representation}. We also show that the 2RC scheme can be derived as a special case of the 3PC.

The proposed scheme is based on block Markov superposition coding to study a general MC with full-duplex bi-directional non-orthogonal cooperation links, unlike~\cite{article14,article19,article21}. It uses short message in order to use DF, as recent schemes tend to do~\cite{article33,unpublished10}. Contrary to~\cite{inproceedings11,unpublished9}, we want to propose a scheme that performs well but that remains manageable. To this end, instead of using respectively PDF as~\cite{inproceedings11}, or partial compress-decode-forward~(PCDF) as~\cite{unpublished9}, which leads to a high number of parameters, we use superpositions of CFs and DFs at each node to obtain a scheme that improves consequently the rate while ensuring a low number of parameters. Note that a superposition of CF and DF is different from a PCDF, and presents a latency advantage that any other rate-splitting strategy~\cite{unpublished10,inproceedings11,unpublished9} does not in general. Indeed, when the PCDF is applied to a short message, one part of the message is decoded with a low latency but the remaining part is decoded only at the end of the scheme due to the symmetry of construction of the scheme, whereas, each layer of the superposition of CF and DF in our proposed scheme is dedicated to a different short message from a different block, which allows an early decoding due to the asymmetry of construction of our scheme. The consequence is that our construction can reduce the latency to 3~blocks for each short message, while in a wide majority of schemes~\cite{inproceedings13,article32,inproceedings11,unpublished9,unpublished10}, all the messages are (fully) decoded after the last block in general.

\subsection{Presentation of the proposed cooperation scheme}
\label{subsec:Presentation of the proposed cooperation scheme}

Suppose without loss of generality that receiver~1 is the first to perform DF, then receiver~2, and finally receiver~3 decodes last, i.e., although all sub-strategies are possible for $k\neq l\neq q$ as illustrated on Fig.~\ref{fig:The MC with receiver cooperation for the 3FC and 3PC schemes - 3FC representation}, we only consider $(l,k,q)=(1,2,3)$. This sub-strategy is denoted $\textrm{STG}_{1,2,3}^{(2,3)}$. We say that a node is ``stronger'' than another one if it decodes any given short message before the other one.

We detail the encoding and decoding related to the short message $m_j,\ j\in[1:b-3]$. In block~$j$, the transmitter sends a codeword as a function of the current message $m_j$ and the past messages $m_{j-2}$ and $m_{j-3}$. The CFs are performed independently, at respectively receiver~2 and~3, to propagate information about $(y_2^n(j),y_3^n(j))$ described by $(k_{j,2},k_{j,3})$ in block~$j$, that are binned into $(l_{j,2},l_{j,3})$ at the end of the block. The bin indices are relayed by the weak nodes toward the stronger nodes in block~$j+1$. In particular, receiver~1 is the destination of all those links, which allows it to perform an early decoding of the message $m_j$ as $\hat{m}_{j,1}$ by jointly decoding its own observation $y_1^n(j)$ of block~$j$ and the bin indices. At this point, receiver~1 cannot refine information about $m_j$ any further since the decoding step has already been performed, thus it will use DF every time information about this message is needed from then on. Receiver~2 also receives the help from receiver~3, and stores this information for later. The DFs propagate information acquired by the strong nodes toward the weaker nodes in blocks~$j+2$ and~$j+3$. In particular, receiver~1 cooperates with the transmitter by using DF toward receiver~2, and by jointly decoding this information with the one stored in block~$j+1$ and its own observation $y_2^n(j)$ of block~$j$, receiver~2 can decode the message $m_j$ as $\hat{m}_{j,2}$. Finally, in block~$j+3$, receivers~1 and~2 cooperate with the transmitter by using DF toward receiver~3 which decodes the message $m_j$ as $\hat{m}_{j,3}$. Note that this last cooperation is the only one involving the cloud center $u^n(\cdot)$ around which $x^n(\cdot\vert\cdot\vert\cdot),\ x_1^n(\cdot\vert\cdot)$ and $x_2^n(\cdot\vert\cdot)$ are generated, and that it is not transmitted as is on the channel. Even though receiver~3 is considered to be the weakest node, this correlated cooperation towards it allows to strongly increase the achievable multicast rate when the cooperation links are strong enough. Note that we used backward decoding at receiver~3 to ease the proof provided in Appx.~\ref{appx:Description of the Proposed Scheme}, then, instead of performing a sliding window decoding of size~4 from block~$j$ to~$j+3$ that would lead to a latency of only $3$~blocks, the receiver~3 decodes all the short messages at the end of the $b$~blocks, and to decode the message $\hat{m}_j$ and $\hat{m}_{j,3}$, it only needs to know $(\hat{m}_{j+1,3},\hat{m}_{j+3,3})$. One can show that $(\hat{m}_{j,1},\hat{m}_{j,2},\hat{m}_{j,3})=(m_j,m_j,m_j)$ with high probability if the rate satisfies Prop.~\ref{prop:3FC}.

Those steps are repeated and superposed as presented in Tab.~\ref{tab:Encoding, transmission, quantization distortion, and decoding for the MC with receiver cooperation for the 3FC scheme} of Appx.~\ref{appx:Description of the Proposed Scheme} for all the short messages, so that any four adjacent blocks are linked together through a block Markov superposition coding scheme, and that those four blocks are necessary and sufficient for all receivers to decode the corresponding message. The CF phase forms a ``all to one structure'' to initiate the scheme quickly and start from the first round the decoding of the current message, whereas the DF phase forms a ``chain structure'' to use a correlated layer in the stack of superposition of the receivers and of the transmitter, toward the next receiver in the chain to improve the achievable rate.

\begin{proposition}[Three-receiver fully interactive cooperation scheme]\label{prop:3FC}
	With the proposed 3FC scheme, we achieve the following lower bound,
	\begin{align}
		C\geq&R_{\text{3FC}}\triangleq\max_{k\neq l\neq q}\bigg\{\max_{\mathcal{P}_{l,k,q}^{(k,q)}}\Big\{\min\Big\{MISO_{3\times1,q}',\nonumber\\
		&\frac{1}{2}(SIMO_{1\times2,l\tilde{k}}+SIMO_{1\times2,l\tilde{q}}+A_1-\mathscr{R}_{k\vert l\tilde{q}}-\mathscr{R}_{q\vert l\tilde{k}}),\nonumber\\
		&\frac{1}{2}(SIMO_{1\times2,l\tilde{q}}+SISO_k+A_2+A_1-\mathscr{R}_{k\vert l\tilde{q}}-\mathscr{R}_{q\vert k}),\nonumber\\
		&SIMO_{1\times2,l\tilde{q}}+A_3-\mathscr{R}_{k\vert l\tilde{q}},SIMO_{1\times2,l\tilde{k}}+A_4-\mathscr{R}_{q\vert l\tilde{k}},\nonumber\\
		&SISO_k+A_2+A_4-\mathscr{R}_{q\vert k},SIMO_{1\times2,l\tilde{k}}+A_5-\mathscr{R}_{q\vert l\tilde{k}},\nonumber\\
		&MISO_{3\times1,k}-\mathscr{R}_{q\vert k},MISO_{2\times1,l}+A_4-\mathscr{R}_{k\vert l}-\mathscr{R}_{q\vert l\tilde{k}},\nonumber\\
		&MISO_{3\times1,l}-\mathscr{R}_{k\vert l}-\mathscr{R}_{q\vert l\tilde{k}},MISO_{2\times1,l}+A_5-\mathscr{R}_{k\vert l}-\mathscr{R}_{q\vert l\tilde{k}},\nonumber\\
		&SIMO_{1\times3,l\tilde{k}\tilde{q}},SIMO_{1\times2,k\tilde{q}}+A_2\Big\}\Big\}\bigg\}\label{eq:3FC}
  \end{align}
	where the MISO three/two to one interference-free terms are,
	\begin{align}
		MISO_{3\times1,l}=&I(X,X_k,X_q;Y_l\vert U,X_l)\\
		MISO_{3\times1,k}=&I(X,X_l,X_q;Y_k\vert U,X_k)\\
		MISO_{3\times1,q}'=&I(U,X,X_k,X_l;Y_q\vert X_q)=I(X,X_k,X_l;Y_q\vert X_q)\\
		MISO_{2\times1,l}=&I(X,X_k;Y_l\vert U,X_l,X_q),
	\end{align}
	the SISO interference-free term is,
	\begin{align}
		SISO_k=I(X;Y_k\vert U,X_l,X_k,X_q),
	\end{align}
	the SIMO one to two/three interference-free terms are,
	\begin{align}
		SIMO_{1\times2,l\tilde{k}}=&I(X;Y_l,\tilde{Y}_k\vert U,X_l,X_k,X_q)\\
		SIMO_{1\times2,l\tilde{q}}=&I(X;Y_l,\tilde{Y}_q\vert U,X_l,X_k,X_q)\\
		SIMO_{1\times2,k\tilde{q}}=&I(X;Y_k,\tilde{Y}_q\vert U,X_l,X_k,X_q)\\
		SIMO_{1\times3,l\tilde{k}\tilde{q}}=&I(X;Y_l,\tilde{Y}_k,\tilde{Y}_q\vert U,X_l,X_k,X_q),
	\end{align}
	the other terms are,
	\begin{align}
		A_1=&I(X_k,X_q;Y_l\vert U,X_l)\\
		A_2=&I(X_l;Y_k\vert U,X_k)\\
		A_3=&I(X_k;Y_l\vert U,X_l,X_q)\\
		A_4=&I(X_q;Y_l\vert U,X_l,X_k)\\
		A_5=&I(X_q;Y_k\vert U,X_l,X_k),
	\end{align}
	the interference-free loss terms induced by the compression are,
	\begin{align}
		\mathscr{R}_{k\vert l}=&I(Y_k;\tilde{Y}_k\vert U,X,X_l,X_k,X_q,Y_l)\\
		\mathscr{R}_{k\vert l\tilde{q}}=&I(Y_k;\tilde{Y}_k\vert U,X,X_l,X_k,X_q,Y_l,\tilde{Y}_q)\\
		\mathscr{R}_{q\vert k}=&I(Y_q;\tilde{Y}_q\vert U,X,X_l,X_k,X_q,Y_k)\\
		\mathscr{R}_{q\vert l\tilde{k}}=&I(Y_q;\tilde{Y}_q\vert U,X,X_l,X_k,X_q,Y_l,\tilde{Y}_k),
	\end{align}
  and where $\mathcal{P}_{l,k,q}^{(k,q)}$ is the set of distributions
\begin{align}
	p(u)p(x_l\vert u)p(x\vert x_l)p(x_k\vert u)p(x_q)p(\tilde{y}_k\vert x_k,y_k)p(\tilde{y}_q\vert x_q,y_q)
\end{align}
with $\vert\mathcal{\tilde{Y}}_k\vert\leq\vert\mathcal{X}_k\vert\vert\mathcal{Y}_k\vert+1$ and $\vert\mathcal{\tilde{Y}}_q\vert\leq\vert\mathcal{X}_q\vert\vert\mathcal{Y}_q\vert+1$. The term with apostrophe $\cdot'$ is the only one involving the random variable $U$ to cooperate toward the weakest node of the completely symmetric cooperation case.
\end{proposition}

We omitted the time-sharing random variable in all the information-theoretic bounds presented in this paper for brevity.

Note that due to the decoding operations and to the perfect CSI, cooperation is possible between the receivers that have already decoded in order to correlate their codewords. In practice this operation, in which separated nodes sharing the same information cooperate to transmit it to another node, is called distributed MIMO or network beamforming~\cite{article39:article40,article36,article42}. It is shown that when the cooperation is implemented correctly, the spatial diversity gain is greater than if it was performed by antennas confined to the same node. For fast-fading channels it is advantageous to use independent inputs so that all nodes can use the same encoder for all channel states.

This scheme only requires superposition of CFs and DFs, both of which are well known. Moreover, self-interference cancellation, full-duplex and distributed cooperation techniques exist to support our concept and continue to be developed. Those bounds can be applied to any given channel compliant to the corresponding requirements, which makes the information-theoretic derivation very interesting.

\subsection{Special cases}
\label{subsec:Special cases}

As a special case of the 3FC, we can get the 3PC by turning off the CF cooperation of the weakest node of the completely symmetric cooperation case, as illustrated in a simplified manner in Fig.~\ref{fig:The MC with receiver cooperation for the 3FC and 3PC schemes - 3PC representation}. Since the 3PC scheme is a special case of the 3FC scheme, we get $R_{\text{3FC}}\geq R_{\text{3PC}}$.

\begin{corollary}[Three-receiver partially interactive cooperation scheme]\label{coro:3PC}
	With the 3PC scheme, we achieve the following lower bound,
	\begin{align}
		C\geq&R_{\text{3PC}}\triangleq\max_{k\neq l\neq q}\bigg\{\max_{\mathcal{P}_{l,k,q}^{(k)}}\Big\{\min\Big\{I(X,X_k,X_l;Y_q\vert X_q),\nonumber\\
		&I(X,X_l;Y_k\vert U,X_k,X_q),I(X;Y_l,\tilde{Y}_k\vert U,X_k,X_l,X_q),\nonumber\\
		&I(X,X_k;Y_l\vert U,X_l,X_q)-I(Y_k;\tilde{Y}_k\vert U,X,X_k,X_l,X_q,Y_l)\Big\}\Big\}\bigg\}
	\end{align}
	where $\mathcal{P}_{l,k,q}^{(k)}$ is the set of distributions $p(u)p(x_l\vert u)p(x\vert x_l)p(x_k\vert u)p(x_q)p(\tilde{y}_k\vert x_k,y_k)$ with $\vert\mathcal{\tilde{Y}}_k\vert\leq\vert\mathcal{X}_k\vert\vert\mathcal{Y}_k\vert+1$.
\end{corollary}

\begin{IEEEproof}
	See Appx.~\ref{appx:Special case: 3PC}
\end{IEEEproof}

As a special case of the 3FC, we can get the 3FC in the Gaussian case. The bounds are further explained in Sec.~\ref{sec:Numerical Results} with the help of Fig.~\ref{fig:Comparison of different cooperation schemes concerning their achievable rate for the Gaussian MC with receiver cooperation}.

\begin{corollary}[3FC Gaussian channel]\label{coro:3FC Gaussian}
	The proposed 3FC scheme achieves the following lower bound expressed explicitly from Prop.~\ref{prop:3FC} in a SISO Gaussian MC
	\begin{align}
		C\geq&R_{\text{3FC}}^{\textrm{Gauss}}\triangleq\max_{k\neq l\neq q}\bigg\{\max_{\mat{\Sigma_{l,k,q}^{(k,q)}}\succeq0,\mat{\Sigma_{l,k,q}^{(k,q)}}\preceq P\mat{I_5}}\bigg\{\min\bigg\{\mathscr{C}(\beta_q'),\nonumber\\
		&\frac{1}{2}\left(\mathscr{C}\left(\gamma_l+\frac{\gamma_k}{1+\Delta_k}\right)+\mathscr{C}\left(\gamma_l+\frac{\gamma_q}{1+\Delta_q}\right)+\mathscr{C}(\beta_l)-\mathscr{C}(\gamma_l)-\mathscr{R}_k-\mathscr{R}_q\right),\nonumber\\
		&\frac{1}{2}\left(\mathscr{C}\left(\gamma_l+\frac{\gamma_q}{1+\Delta_q}\right)+\mathscr{C}(\gamma_k)+\mathscr{C}(\beta_k)-\mathscr{C}(\kappa_k)+\mathscr{C}(\beta_l)-\mathscr{C}(\gamma_l)-\mathscr{R}_k-\mathscr{R}_q\right),\nonumber\\
		&\mathscr{C}\left(\gamma_l+\frac{\gamma_q}{1+\Delta_q}\right)+\mathscr{C}(\kappa_l)-\mathscr{C}(\gamma_l)-\mathscr{R}_k,\mathscr{C}\left(\gamma_l+\frac{\gamma_k}{1+\Delta_k}\right)+\mathscr{C}(\lambda_l)-\mathscr{C}(\gamma_l)-\mathscr{R}_q,\nonumber\\
		&\mathscr{C}(\gamma_k)+\mathscr{C}(\beta_k)-\mathscr{C}(\kappa_k)+\mathscr{C}(\lambda_l)-\mathscr{C}(\gamma_l)-\mathscr{R}_q,\nonumber\\
		&\mathscr{C}\left(\gamma_l+\frac{\gamma_k}{1+\Delta_k}\right)+\mathscr{C}(\kappa_k)-\mathscr{C}(\gamma_k)-\mathscr{R}_q,\nonumber\\
		&\mathscr{C}(\beta_k)-\mathscr{R}_q,\mathscr{C}(\kappa_l)+\mathscr{C}(\lambda_l)-\mathscr{C}(\gamma_l)-\mathscr{R}_k-\mathscr{R}_q,\nonumber\\
		&\mathscr{C}(\beta_l)-\mathscr{R}_k-\mathscr{R}_q,\mathscr{C}(\kappa_l)+\mathscr{C}(\kappa_k)-\mathscr{C}(\gamma_k)-\mathscr{R}_k-\mathscr{R}_q,\nonumber\\
		&\mathscr{C}\left(\gamma_l+\frac{\gamma_k}{1+\Delta_k}+\frac{\gamma_q}{1+\Delta_q}\right),\mathscr{C}\left(\gamma_k+\frac{\gamma_q}{1+\Delta_q}\right)+\mathscr{C}(\beta_k)-\mathscr{C}(\kappa_k)\bigg\}\bigg\}\bigg\}
	\end{align}
	where the terms corresponding to MISO three to one interference-free mutual information terms are composed of,
	\begin{align}
		\beta_l=&\textrm{SNR}_l\rho_{X'}+\textrm{SNR}_{kl}\rho_{X_k'}+\textrm{SNR}_{ql}\rho_{X_q}\\
		\beta_k=&\textrm{SNR}_k(\rho_{X'}+\rho_{X_l'}\rho_{A_k}^2)+\textrm{SNR}_{lk}\rho_{X_l'}+\textrm{SNR}_{qk}\rho_{X_q}+\nonumber\\
		&2\sqrt{\textrm{SNR}_k\textrm{SNR}_{lk}}\rho_{X_l'}\rho_{A_k}\cos(\theta_{A_k})\\
		\beta_q'=&\textrm{SNR}_q(\rho_{X'}+\rho_{X_l'}\rho_{A_k}^2+\rho_U\rho_{A_l}^2\rho_{A_k}^2)+\textrm{SNR}_{lq}(\rho_{X_l'}+\rho_U\rho_{A_l}^2)+\textrm{SNR}_{kq}(\rho_{X_k'}+\rho_U\rho_{B_l}^2)+\nonumber\\
		&2\sqrt{\textrm{SNR}_q\textrm{SNR}_{lq}}(\rho_{X_l'}\rho_{A_k}+\rho_U\rho_{A_l}^2\rho_{A_k})\cos(\theta_{A_k})+\nonumber\\
		&2\sqrt{\textrm{SNR}_q\textrm{SNR}_{kq}}\rho_U\rho_{A_l}\rho_{A_k}\rho_{B_l}\cos(\theta_{A_l}+\theta_{A_k}-\theta_{B_l})+\nonumber\\
		&2\sqrt{\textrm{SNR}_{lq}\textrm{SNR}_{kq}}\rho_U\rho_{A_l}\rho_{B_l}\cos(\theta_{A_l}-\theta_{B_l}),
	\end{align}
	the terms corresponding to MISO two to one interference-free mutual information terms are composed of,
	\begin{align}
		\kappa_l=&\textrm{SNR}_l\rho_{X'}+\textrm{SNR}_{kl}\rho_{X_k'}\\
		\lambda_l=&\textrm{SNR}_l\rho_{X'}+\textrm{SNR}_{ql}\rho_{X_q}\\
		\kappa_k=&\textrm{SNR}_k\rho_{X'}+\textrm{SNR}_{qk}\rho_{X_q},
	\end{align}
	the terms corresponding to SISO interference-free mutual information terms are composed of,
	\begin{align}
		\gamma_l=&\textrm{SNR}_l\rho_{X'}\\
		\gamma_k=&\textrm{SNR}_k\rho_{X'}\\
		\gamma_q=&\textrm{SNR}_q\rho_{X'},
	\end{align}
	the interference-free loss terms induced by the compression are,
	\begin{align}
		\mathscr{R}_k=&\mathscr{C}\left(\frac{1}{\Delta_k}\right)\\
		\mathscr{R}_q=&\mathscr{C}\left(\frac{1}{\Delta_q}\right),
	\end{align}
	and where the subscripts in $\beta_\cdot,\ \kappa_\cdot,\ \lambda_\cdot,\ \gamma_\cdot$ correspond to the destination index. The covariance matrix $\mat{\Sigma_{l,k,q}^{(k,q)}}$ is defined in~\eqref{eq:covariance matrix} of Appx.~\ref{appx:Special case: 3FC in the Gaussian case}, and includes the correlation coefficients $0\leq\rho_U,\rho_{X_l'},\rho_{A_l},\rho_{X'},\rho_{A_k},\rho_{X_k'},\rho_{B_l},\rho_{X_q}\leq1,\ \theta_{A_l},\theta_{A_k},\theta_{B_l}\in[0,2\pi)$. The compression noise powers are $0\leq\Delta_k,\Delta_q$.
\end{corollary}

\begin{IEEEproof}
	See Appx.~\ref{appx:Special case: 3FC in the Gaussian case}
\end{IEEEproof}

As a special case of the 3PC, one can get the 2RC~\cite{ownpublications1} by not requiring further the weakest receiver of the completely symmetric cooperation case to decode anymore.

\begin{IEEEproof}
	See Appx.~\ref{appx:Special case: 2RC}
\end{IEEEproof}

\section{Numerical Results}
\label{sec:Numerical Results}

\begin{figure*}[t]
	\begin{subfigure}[t]{0.325\textwidth}
		\begin{tikzpicture}[line width=.75pt,font=\footnotesize,scale=0.72,every node/.style={scale=0.72}]
			\begin{axis}
				[xmin=-20.0,xmax=30.0,ymin=3.4,ymax=5,xlabel={$\textsf{SNR}_{\textrm{coop}}$~($\si{\decibel}$)},ylabel={Rate~(\si{\bit/\second/\hertz})},grid=major,mark options=solid,mark repeat=3,smooth,tension=0.3,legend pos=south east,ylabel style={yshift=-0.4cm},legend cell align=left];
				\addplot[color=green,dashed,mark=o] coordinates{(-20.0,3.459)(-19.0,3.459)(-18.0,3.459)(-17.0,3.459)(-16.0,3.459)(-15.0,3.459)(-14.0,3.459)(-13.0,3.459)(-12.0,3.459)(-11.0,3.459)(-10.0,3.459)(-9.0,3.459)(-8.0,3.459)(-7.0,3.459)(-6.0,3.459)(-5.0,3.459)(-4.0,3.459)(-3.0,3.459)(-2.0,3.459)(-1.0,3.459)(0.0,3.459)(1.0,3.459)(2.0,3.459)(3.0,3.459)(4.0,3.459)(5.0,3.459)(6.0,3.459)(7.0,3.459)(8.0,3.459)(9.0,3.459)(10.0,3.459)(11.0,3.459)(12.0,3.459)(13.0,3.459)(14.0,3.459)(15.0,3.459)(16.0,3.459)(17.0,3.459)(18.0,3.459)(19.0,3.459)(20.0,3.459)(21.0,3.459)(22.0,3.459)(23.0,3.459)(24.0,3.459)(25.0,3.459)(26.0,3.459)(27.0,3.459)(28.0,3.459)(29.0,3.459)(30.0,3.459)};
				\addplot[color=magenta,dashed,mark=square] coordinates{(-20.0,3.467)(-19.0,3.468)(-18.0,3.470)(-17.0,3.472)(-16.0,3.475)(-15.0,3.479)(-14.0,3.483)(-13.0,3.489)(-12.0,3.496)(-11.0,3.505)(-10.0,3.519)(-9.0,3.540)(-8.0,3.560)(-7.0,3.588)(-6.0,3.616)(-5.0,3.663)(-4.0,3.713)(-3.0,3.774)(-2.0,3.830)(-1.0,3.888)(0.0,3.958)(1.0,4.036)(2.0,4.133)(3.0,4.214)(4.0,4.297)(5.0,4.407)(6.0,4.521)(7.0,4.629)(8.0,4.740)(9.0,4.841)(10.0,4.954)(11.0,4.954)(12.0,4.954)(13.0,4.954)(14.0,4.954)(15.0,4.954)(16.0,4.954)(17.0,4.954)(18.0,4.954)(19.0,4.954)(20.0,4.954)(21.0,4.954)(22.0,4.954)(23.0,4.954)(24.0,4.954)(25.0,4.954)(26.0,4.954)(27.0,4.954)(28.0,4.954)(29.0,4.954)(30.0,4.954)};
				\addplot[color=red,mark=diamond] coordinates{(-20.0,3.461)(-19.0,3.461)(-18.0,3.461)(-17.0,3.462)(-16.0,3.462)(-15.0,3.463)(-14.0,3.464)(-13.0,3.465)(-12.0,3.467)(-11.0,3.469)(-10.0,3.471)(-9.0,3.474)(-8.0,3.478)(-7.0,3.484)(-6.0,3.490)(-5.0,3.498)(-4.0,3.508)(-3.0,3.520)(-2.0,3.532)(-1.0,3.547)(0.0,3.564)(1.0,3.585)(2.0,3.614)(3.0,3.651)(4.0,3.693)(5.0,3.740)(6.0,3.789)(7.0,3.841)(8.0,3.898)(9.0,3.961)(10.0,4.026)(11.0,4.087)(12.0,4.149)(13.0,4.215)(14.0,4.276)(15.0,4.338)(16.0,4.395)(17.0,4.444)(18.0,4.494)(19.0,4.546)(20.0,4.594)(21.0,4.634)(22.0,4.671)(23.0,4.703)(24.0,4.727)(25.0,4.749)(26.0,4.771)(27.0,4.789)(28.0,4.809)(29.0,4.825)(30.0,4.839)};
				\addplot[color=blue,mark=triangle] coordinates{(-20.0,3.460)(-19.0,3.460)(-18.0,3.460)(-17.0,3.461)(-16.0,3.461)(-15.0,3.461)(-14.0,3.462)(-13.0,3.463)(-12.0,3.463)(-11.0,3.464)(-10.0,3.466)(-9.0,3.467)(-8.0,3.469)(-7.0,3.472)(-6.0,3.475)(-5.0,3.479)(-4.0,3.484)(-3.0,3.490)(-2.0,3.497)(-1.0,3.506)(0.0,3.518)(1.0,3.532)(2.0,3.549)(3.0,3.569)(4.0,3.593)(5.0,3.622)(6.0,3.655)(7.0,3.692)(8.0,3.734)(9.0,3.781)(10.0,3.830)(11.0,3.882)(12.0,3.936)(13.0,3.988)(14.0,4.040)(15.0,4.088)(16.0,4.133)(17.0,4.174)(18.0,4.210)(19.0,4.241)(20.0,4.268)(21.0,4.291)(22.0,4.310)(23.0,4.325)(24.0,4.338)(25.0,4.349)(26.0,4.357)(27.0,4.364)(28.0,4.370)(29.0,4.375)(30.0,4.378)};
				\addplot[color=black,mark=x] coordinates{(-20.0,3.461)(-19.0,3.461)(-18.0,3.461)(-17.0,3.462)(-16.0,3.462)(-15.0,3.463)(-14.0,3.464)(-13.0,3.465)(-12.0,3.467)(-11.0,3.469)(-10.0,3.471)(-9.0,3.474)(-8.0,3.478)(-7.0,3.484)(-6.0,3.490)(-5.0,3.498)(-4.0,3.508)(-3.0,3.520)(-2.0,3.532)(-1.0,3.547)(0.0,3.564)(1.0,3.585)(2.0,3.614)(3.0,3.651)(4.0,3.693)(5.0,3.740)(6.0,3.789)(7.0,3.841)(8.0,3.898)(9.0,3.961)(10.0,4.026)(11.0,4.087)(12.0,4.149)(13.0,4.215)(14.0,4.276)(15.0,4.338)(16.0,4.395)(17.0,4.444)(18.0,4.494)(19.0,4.546)(20.0,4.594)(21.0,4.634)(22.0,4.671)(23.0,4.703)(24.0,4.727)(25.0,4.749)(26.0,4.771)(27.0,4.789)(28.0,4.809)(29.0,4.825)(30.0,4.839)};
				\legend{{No Cooperation},{Cutset Upper Bound},{Noisy Network Coding~(NNC)},{3PC},{Proposed Scheme~(3FC)}};
			\end{axis}
		\end{tikzpicture}
		\caption{$\textsf{SNR}_1=\textsf{SNR}_2=\textsf{SNR}_3=\SI{10}{\decibel}$~(symmetric). $R_{\textrm{3FC}}=R_{\textrm{NNC}}$ outperforms $R_{\textrm{3PC}}$ and $R_{\textrm{NC}}$.}
		\label{fig:SISO 10-10-10}
	\end{subfigure}
	~
	\begin{subfigure}[t]{0.31\textwidth}
		\begin{tikzpicture}[line width=.75pt,font=\footnotesize,scale=0.72,every node/.style={scale=0.72}]
			\begin{axis}
				[xmin=-20.0,xmax=30.0,ymin=2.0,ymax=4.4,xlabel={$\textsf{SNR}_{\textrm{coop}}$~($\si{\decibel}$)},grid=major,mark options=solid,mark repeat=3,smooth,tension=0.3];
				\addplot[color=green,dashed,mark=o] coordinates{(-20.0,2.057)(-19.0,2.057)(-18.0,2.057)(-17.0,2.057)(-16.0,2.057)(-15.0,2.057)(-14.0,2.057)(-13.0,2.057)(-12.0,2.057)(-11.0,2.057)(-10.0,2.057)(-9.0,2.057)(-8.0,2.057)(-7.0,2.057)(-6.0,2.057)(-5.0,2.057)(-4.0,2.057)(-3.0,2.057)(-2.0,2.057)(-1.0,2.057)(0.0,2.057)(1.0,2.057)(2.0,2.057)(3.0,2.057)(4.0,2.057)(5.0,2.057)(6.0,2.057)(7.0,2.057)(8.0,2.057)(9.0,2.057)(10.0,2.057)(11.0,2.057)(12.0,2.057)(13.0,2.057)(14.0,2.057)(15.0,2.057)(16.0,2.057)(17.0,2.057)(18.0,2.057)(19.0,2.057)(20.0,2.057)(21.0,2.057)(22.0,2.057)(23.0,2.057)(24.0,2.057)(25.0,2.057)(26.0,2.057)(27.0,2.057)(28.0,2.057)(29.0,2.057)(30.0,2.057)};
				\addplot[color=magenta,dashed,mark=square] coordinates{(-20.0,2.215)(-19.0,2.235)(-18.0,2.257)(-17.0,2.282)(-16.0,2.310)(-15.0,2.342)(-14.0,2.377)(-13.0,2.414)(-12.0,2.452)(-11.0,2.497)(-10.0,2.547)(-9.0,2.601)(-8.0,2.659)(-7.0,2.721)(-6.0,2.789)(-5.0,2.869)(-4.0,2.945)(-3.0,3.035)(-2.0,3.120)(-1.0,3.228)(0.0,3.329)(1.0,3.461)(2.0,3.575)(3.0,3.701)(4.0,3.833)(5.0,3.969)(6.0,4.112)(7.0,4.210)(8.0,4.261)(9.0,4.261)(10.0,4.261)(11.0,4.261)(12.0,4.261)(13.0,4.261)(14.0,4.261)(15.0,4.261)(16.0,4.261)(17.0,4.261)(18.0,4.261)(19.0,4.261)(20.0,4.261)(21.0,4.261)(22.0,4.261)(23.0,4.261)(24.0,4.261)(25.0,4.261)(26.0,4.261)(27.0,4.261)(28.0,4.261)(29.0,4.261)(30.0,4.261)};
				\addplot[color=red,mark=diamond] coordinates{(-20.0,2.061)(-19.0,2.062)(-18.0,2.063)(-17.0,2.065)(-16.0,2.067)(-15.0,2.070)(-14.0,2.073)(-13.0,2.077)(-12.0,2.081)(-11.0,2.088)(-10.0,2.094)(-9.0,2.104)(-8.0,2.117)(-7.0,2.134)(-6.0,2.153)(-5.0,2.170)(-4.0,2.197)(-3.0,2.231)(-2.0,2.267)(-1.0,2.324)(0.0,2.377)(1.0,2.443)(2.0,2.508)(3.0,2.575)(4.0,2.651)(5.0,2.762)(6.0,2.856)(7.0,2.960)(8.0,3.047)(9.0,3.150)(10.0,3.246)(11.0,3.379)(12.0,3.470)(13.0,3.543)(14.0,3.616)(15.0,3.680)(16.0,3.742)(17.0,3.803)(18.0,3.844)(19.0,3.888)(20.0,3.920)(21.0,3.950)(22.0,3.980)(23.0,4.007)(24.0,4.034)(25.0,4.059)(26.0,4.082)(27.0,4.099)(28.0,4.122)(29.0,4.138)(30.0,4.154)};
				\addplot[color=blue,mark=triangle] coordinates{(-20.0,2.064)(-19.0,2.066)(-18.0,2.068)(-17.0,2.071)(-16.0,2.075)(-15.0,2.079)(-14.0,2.085)(-13.0,2.092)(-12.0,2.100)(-11.0,2.111)(-10.0,2.125)(-9.0,2.142)(-8.0,2.163)(-7.0,2.189)(-6.0,2.222)(-5.0,2.261)(-4.0,2.310)(-3.0,2.369)(-2.0,2.439)(-1.0,2.524)(0.0,2.623)(1.0,2.735)(2.0,2.850)(3.0,2.969)(4.0,3.065)(5.0,3.195)(6.0,3.319)(7.0,3.460)(8.0,3.620)(9.0,3.662)(10.0,3.692)(11.0,3.723)(12.0,3.754)(13.0,3.785)(14.0,3.814)(15.0,3.841)(16.0,3.865)(17.0,3.887)(18.0,3.907)(19.0,3.923)(20.0,3.937)(21.0,3.949)(22.0,3.959)(23.0,3.967)(24.0,3.974)(25.0,3.979)(26.0,3.984)(27.0,3.987)(28.0,3.990)(29.0,3.992)(30.0,3.994)};
				\addplot[color=black,mark=x] coordinates{(-20.0,2.064)(-19.0,2.066)(-18.0,2.068)(-17.0,2.071)(-16.0,2.075)(-15.0,2.079)(-14.0,2.085)(-13.0,2.092)(-12.0,2.100)(-11.0,2.111)(-10.0,2.125)(-9.0,2.142)(-8.0,2.163)(-7.0,2.189)(-6.0,2.222)(-5.0,2.261)(-4.0,2.310)(-3.0,2.369)(-2.0,2.439)(-1.0,2.524)(0.0,2.623)(1.0,2.735)(2.0,2.850)(3.0,2.970)(4.0,3.066)(5.0,3.196)(6.0,3.321)(7.0,3.473)(8.0,3.623)(9.0,3.678)(10.0,3.725)(11.0,3.767)(12.0,3.803)(13.0,3.839)(14.0,3.882)(15.0,3.918)(16.0,3.949)(17.0,3.978)(18.0,4.007)(19.0,4.030)(20.0,4.053)(21.0,4.070)(22.0,4.090)(23.0,4.105)(24.0,4.124)(25.0,4.143)(26.0,4.154)(27.0,4.170)(28.0,4.179)(29.0,4.188)(30.0,4.196)};
			\end{axis}
		\end{tikzpicture}
		\caption{$\textsf{SNR}_1=\SI{10}{\decibel},\ \textsf{SNR}_2=\SI{7}{\decibel},\ \textsf{SNR}_3=\SI{5}{\decibel}$~(asymmetric). $R_{\textrm{3FC}}$ outperforms $R_{\textrm{3PC}},\ R_{\textrm{NNC}}$ and $R_{\textrm{NC}}$.}
		\label{fig:SISO 10-7-5}
	\end{subfigure}
	~
	\begin{subfigure}[t]{0.31\textwidth}
		\begin{tikzpicture}[line width=.75pt,font=\footnotesize,scale=0.72,every node/.style={scale=0.72}]
			\begin{axis}
				[xmin=-20.0,xmax=30.0,ymin=0.9,ymax=4,xlabel={$\textsf{SNR}_{\textrm{coop}}$~($\si{\decibel}$)},grid=major,mark options=solid,mark repeat=3,smooth,tension=0.3];
				\addplot[color=green,dashed,mark=o] coordinates{(-20.0,1.000)(-19.0,1.000)(-18.0,1.000)(-17.0,1.000)(-16.0,1.000)(-15.0,1.000)(-14.0,1.000)(-13.0,1.000)(-12.0,1.000)(-11.0,1.000)(-10.0,1.000)(-9.0,1.000)(-8.0,1.000)(-7.0,1.000)(-6.0,1.000)(-5.0,1.000)(-4.0,1.000)(-3.0,1.000)(-2.0,1.000)(-1.0,1.000)(0.0,1.000)(1.0,1.000)(2.0,1.000)(3.0,1.000)(4.0,1.000)(5.0,1.000)(6.0,1.000)(7.0,1.000)(8.0,1.000)(9.0,1.000)(10.0,1.000)(11.0,1.000)(12.0,1.000)(13.0,1.000)(14.0,1.000)(15.0,1.000)(16.0,1.000)(17.0,1.000)(18.0,1.000)(19.0,1.000)(20.0,1.000)(21.0,1.000)(22.0,1.000)(23.0,1.000)(24.0,1.000)(25.0,1.000)(26.0,1.000)(27.0,1.000)(28.0,1.000)(29.0,1.000)(30.0,1.000)};
				\addplot[color=magenta,dashed,mark=square] coordinates{(-20.0,1.203)(-19.0,1.229)(-18.0,1.258)(-17.0,1.291)(-16.0,1.329)(-15.0,1.371)(-14.0,1.419)(-13.0,1.473)(-12.0,1.533)(-11.0,1.601)(-10.0,1.667)(-9.0,1.740)(-8.0,1.810)(-7.0,1.885)(-6.0,1.980)(-5.0,2.096)(-4.0,2.208)(-3.0,2.335)(-2.0,2.454)(-1.0,2.600)(0.0,2.736)(1.0,2.871)(2.0,3.021)(3.0,3.170)(4.0,3.344)(5.0,3.515)(6.0,3.663)(7.0,3.800)(8.0,3.922)(9.0,3.922)(10.0,3.922)(11.0,3.922)(12.0,3.922)(13.0,3.922)(14.0,3.922)(15.0,3.922)(16.0,3.922)(17.0,3.922)(18.0,3.922)(19.0,3.922)(20.0,3.922)(21.0,3.922)(22.0,3.922)(23.0,3.922)(24.0,3.922)(25.0,3.922)(26.0,3.922)(27.0,3.922)(28.0,3.922)(29.0,3.922)(30.0,3.922)};
				\addplot[color=red,mark=diamond] coordinates{(-20.0,1.010)(-19.0,1.013)(-18.0,1.016)(-17.0,1.020)(-16.0,1.025)(-15.0,1.031)(-14.0,1.038)(-13.0,1.048)(-12.0,1.061)(-11.0,1.078)(-10.0,1.098)(-9.0,1.122)(-8.0,1.149)(-7.0,1.181)(-6.0,1.220)(-5.0,1.268)(-4.0,1.330)(-3.0,1.402)(-2.0,1.483)(-1.0,1.571)(0.0,1.666)(1.0,1.775)(2.0,1.896)(3.0,2.021)(4.0,2.160)(5.0,2.297)(6.0,2.431)(7.0,2.574)(8.0,2.716)(9.0,2.841)(10.0,2.958)(11.0,3.048)(12.0,3.145)(13.0,3.235)(14.0,3.320)(15.0,3.393)(16.0,3.455)(17.0,3.507)(18.0,3.550)(19.0,3.585)(20.0,3.620)(21.0,3.650)(22.0,3.681)(23.0,3.706)(24.0,3.727)(25.0,3.752)(26.0,3.770)(27.0,3.786)(28.0,3.800)(29.0,3.816)(30.0,3.826)};
				\addplot[color=blue,mark=triangle] coordinates{(-20.0,1.014)(-19.0,1.018)(-18.0,1.023)(-17.0,1.029)(-16.0,1.036)(-15.0,1.045)(-14.0,1.056)(-13.0,1.071)(-12.0,1.088)(-11.0,1.110)(-10.0,1.138)(-9.0,1.171)(-8.0,1.212)(-7.0,1.262)(-6.0,1.323)(-5.0,1.396)(-4.0,1.483)(-3.0,1.586)(-2.0,1.706)(-1.0,1.843)(0.0,2.000)(1.0,2.171)(2.0,2.348)(3.0,2.530)(4.0,2.695)(5.0,2.860)(6.0,3.027)(7.0,3.195)(8.0,3.386)(9.0,3.586)(10.0,3.611)(11.0,3.629)(12.0,3.651)(13.0,3.672)(14.0,3.695)(15.0,3.716)(16.0,3.735)(17.0,3.749)(18.0,3.763)(19.0,3.774)(20.0,3.783)(21.0,3.791)(22.0,3.797)(23.0,3.802)(24.0,3.807)(25.0,3.810)(26.0,3.813)(27.0,3.815)(28.0,3.817)(29.0,3.818)(30.0,3.819)};
				\addplot[color=black,mark=x] coordinates{(-20.0,1.014)(-19.0,1.018)(-18.0,1.023)(-17.0,1.029)(-16.0,1.036)(-15.0,1.045)(-14.0,1.056)(-13.0,1.071)(-12.0,1.088)(-11.0,1.110)(-10.0,1.138)(-9.0,1.171)(-8.0,1.212)(-7.0,1.262)(-6.0,1.323)(-5.0,1.396)(-4.0,1.483)(-3.0,1.586)(-2.0,1.706)(-1.0,1.843)(0.0,2.000)(1.0,2.171)(2.0,2.348)(3.0,2.530)(4.0,2.695)(5.0,2.860)(6.0,3.030)(7.0,3.198)(8.0,3.389)(9.0,3.594)(10.0,3.625)(11.0,3.648)(12.0,3.673)(13.0,3.690)(14.0,3.714)(15.0,3.736)(16.0,3.755)(17.0,3.770)(18.0,3.786)(19.0,3.799)(20.0,3.811)(21.0,3.823)(22.0,3.833)(23.0,3.842)(24.0,3.850)(25.0,3.858)(26.0,3.865)(27.0,3.871)(28.0,3.875)(29.0,3.882)(30.0,3.886)};
			\end{axis}
		\end{tikzpicture}
		\caption{$\textsf{SNR}_1=\SI{10}{\decibel},\ \textsf{SNR}_2=\SI{5}{\decibel},\ \textsf{SNR}_3=\SI{0}{\decibel}$~(asymmetric). $R_{\textrm{3FC}}$ outperforms $R_{\textrm{3PC}},\ R_{\textrm{NNC}}$ and $R_{\textrm{NC}}$.}
		\label{fig:SISO 10-5-0}
	\end{subfigure}
	\caption{Comparison of different cooperation schemes concerning their achievable rate for the Gaussian MC with receiver cooperation.}
	\label{fig:Comparison of different cooperation schemes concerning their achievable rate for the Gaussian MC with receiver cooperation}
\end{figure*}
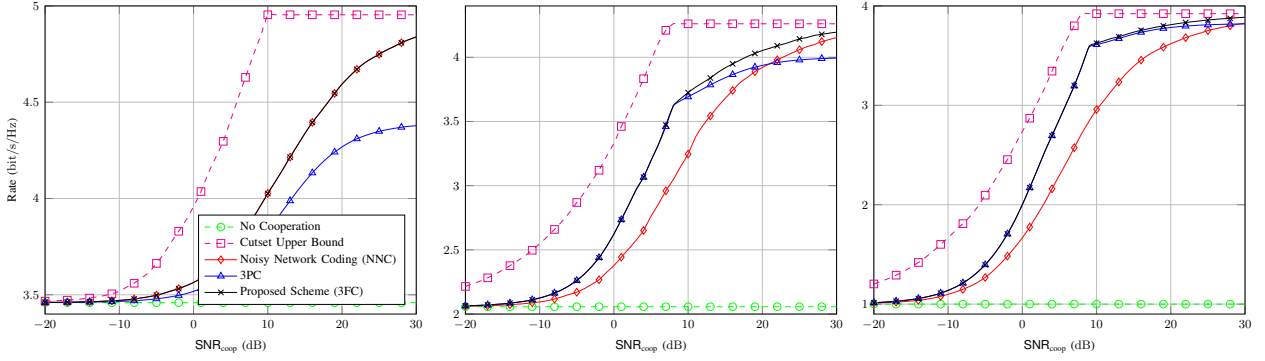

In this section, we focus on the SISO Gaussian MC as defined in~\eqref{eq:AWGN k}, and evaluate through numerical simulations the achievable rate of the proposed scheme given in Prop.~\ref{prop:3FC}, as well as the cutset upper bound~\cite[Th.~18.1]{book3} and three lower bounds: the ``no cooperation" scheme in which the weakest user set the rate, the NNC scheme~\cite[Th.~1]{article32},\cite[Th.~1]{article33}, and the 3PC scheme given in Coro.~\ref{coro:3PC}. Note that to provide a fair comparison the parameters such as input correlation and compression noise variance are optimized for each bound. We study the impact of the cooperation link on the throughput of the channel. We assume that the SNR of the cooperation links is symmetric, i.e., $\textsf{SNR}_{kl}=\textsf{SNR}_{lk}=\textsf{SNR}_{\text{coop}},\ k\neq l,\ (k,l)\in[1:3]^2$, and consider that $\textsf{SNR}_1\geq\textsf{SNR}_2\geq\textsf{SNR}_3$. In Fig.~\ref{fig:Comparison of different cooperation schemes concerning their achievable rate for the Gaussian MC with receiver cooperation}, we fix the SNR of the main channel, and plot the throughput in terms of spectral efficiency~(\si{\bit/\second/\hertz}) by varying $\textsf{SNR}_{\text{coop}}$ from~\SIrange{-20}{30}{\decibel}. In Fig.~\ref{fig:SISO 10-10-10}, the main channel is symmetric with a $\textsf{SNR}$ of $\SI{10}{\decibel}$ at each receiver, while in Fig.~\ref{fig:SISO 10-7-5} and Fig.~\ref{fig:SISO 10-5-0} the main channel is asymmetric. In all cases, both the NNC scheme and the 3FC scheme go from the ``no cooperation'' lower bound $R_{\text{NC}}$ when the cooperation link is weak, to the cutset upper bound $R_{\text{CS}}$ when the cooperation link is strong. The 3PC scheme also grows from the $R_{\text{NC}}$ at low $\textsf{SNR}_{\text{coop}}$ but does not reach the $R_{\text{CS}}$ at high $\textsf{SNR}_{\text{coop}}$. The proposed 3FC scheme outperforms both the NNC and the 3PC schemes in the Gaussian case, thus it is a good generalization of the 2RC scheme.

In Fig.~\ref{fig:SISO 10-7-5}, the $R_{\text{NC}}$ remains at $\mathscr{C}(10^{\frac{\textsf{SNR}_3}{10}})\approx$\SI{2.057}{\bit/\second/\hertz}, while the $R_{\text{CS}}$ goes from the $R_{\text{NC}}$ to the broadcast bottleneck $\mathscr{C}(10^{\frac{\textsf{SNR}_1}{10}}+10^{\frac{\textsf{SNR}_2}{10}}+10^{\frac{\textsf{SNR}_3}{10}})\approx$\SI{4.261}{\bit/\second/\hertz} as the strength of the cooperation link increases. At low $\textsf{SNR}_{\text{coop}}$, the 3FC scheme selects the bound $I(X,X_1,X_2;Y_3\vert X_3)=\mathscr{C}(\beta_3')$ which is squeezed below by the $R_{\text{3PC}}$ (the bound of the 3PC is equivalent in terms of mutual information and the probability distribution is less general), and above by the $R_{\text{CS}}$ (the bound of the cutset upper bound is equivalent in terms of mutual information, but the probability distribution is more general) as the strength of the cooperation link decreases. This bound represents the cooperation of receivers~1, 2, and the transmitter using DF towards receiver~3, and shows that correlating the codewords of the receivers is very helpful when the cooperation link is weak. At high $\textsf{SNR}_{\text{coop}}$, the 3FC scheme selects the bound $I(X;Y_1,\tilde{Y}_2,\tilde{Y}_3\vert U,X_1,X_2,X_3)=\mathscr{C}\left(\gamma_1+\frac{\gamma_2}{1+\Delta_2}+\frac{\gamma_3}{1+\Delta_3}\right)$ which is squeezed below by the $R_{\text{NNC}}$ (the bound of the NNC is equivalent in terms of mutual information, but since the probability distribution is different because of the lack of correlation, it is only equal to $\mathscr{C}\left(\textsf{SNR}_1+\frac{\textsf{SNR}_2}{1+\Delta_2}+\frac{\textsf{SNR}_3}{1+\Delta_3}\right)$ in the Gaussian case), and above by the $R_{\text{CS}}$ (the bound of the cutset upper bound in terms of mutual information is $I(X;Y_1,Y_2,Y_3|X_1,X_2,X_3)$ and the probability distribution is more general) as the strength of the cooperation link increases. This bound represents the broadcast bottleneck, and shows that the cooperation links have to be designed such that every receiver can access all the information of the other receivers when the cooperation links are strong enough, and that once again correlating the codewords of the receivers is very helpful. Note that with a single CF from receivers~2 and~3, the structure of the bound at high $\textsf{SNR}_{\text{coop}}$ is already equivalent to the one of the NNC in terms of mutual information, so there is no need to perform more CF on a short message, as it is also shown in~\cite{article33,ownpublications1}. Note that the 3PC scheme outperforms the NNC scheme with weak cooperation, and conversely with strong cooperation since the active bound at high $\textsf{SNR}_{\text{coop}}$ is only $I(X;Y_1,\tilde{Y}_2|U,X_1,X_2,X_3)$. This leads to the observation that the 3PC scheme remains lower than the 3FC scheme and goes to $\mathscr{C}(10^{\frac{\textsf{SNR}_1}{10}}+10^{\frac{\textsf{SNR}_2}{10}})\approx$\SI{4.001}{\bit/\second/\hertz} since there is no CF link coming from receiver~3, i.e., the information does not flow properly through each node. The comments of Fig.~\ref{fig:SISO 10-7-5} also hold for Fig.~\ref{fig:SISO 10-5-0}. Thus, the gain of the 3FC from $\textsf{SNR}_{\text{coop}}=0$ to $\textsf{SNR}_{\text{coop}}\rightarrow\infty$ is
\begin{align}
	G_{\text{3FC}}=&\log\left(1+\sum_{k=1}^3\textsf{SNR}_k\right)-\log(1+\textsf{SNR}_3)\\
	=&\log\left(1+\frac{\textsf{SNR}_1+\textsf{SNR}_2}{1+\textsf{SNR}_3}\right),\label{eq:3FC gain}
\end{align}
while the gain of the 3PC is
\begin{align}
	G_{\text{3PC}}=\log\left(\frac{1+\textsf{SNR}_1+\textsf{SNR}_2}{1+\textsf{SNR}_3}\right).\label{eq:3PC gain}
\end{align}

In Fig.~\ref{fig:SISO 10-10-10}, the $R_{\text{3FC}}$ is equal to the $R_{\text{NNC}}$. At low $\textsf{SNR}_{\text{coop}}$, the 3FC scheme selects the bound $\mathscr{C}(\beta_1)-\mathscr{R}_2-\mathscr{R}_3$, and the NNC scheme selects the bound $\mathscr{C}(\textrm{SNR}_1+\textrm{SNR}_{21}+\textrm{SNR}_{31})-\mathscr{R}_2-\mathscr{R}_3$. They turn out to be equal in the symmetric case since all receivers achieve the same performance, so the DF operation does not bring any gain, thus $\rho_U=\rho_{X_1'}=\rho_{X'}=\rho_{X_2'}=\rho_{X_3}=1,\ \rho_{A_1}=\rho_{A_2}=\rho_{B_1}=0$ and $\theta_{A_1}=\theta_{A_2}=\theta_{B_1}=0$. At high $\textsf{SNR}_{\text{coop}}$, the 3FC scheme selects the bound $\mathscr{C}\left(\gamma_1+\frac{\gamma_2}{1+\Delta_2}+\frac{\gamma_3}{1+\Delta_3}\right)$, and the NNC scheme selects the bound $\mathscr{C}\left(\textrm{SNR}_1+\frac{\textrm{SNR}_2}{1+\Delta_2}+\frac{\textrm{SNR}_3}{1+\Delta_3}\right)$. They turn out to be equal in the symmetric case for the same reason. In conclusion, at low $\textsf{SNR}_{\text{coop}}$ the 3FC bounds corresponding to the CF are loose, while at high $\textsf{SNR}_{\text{coop}}$ the DF ones are loose. In the middle range of $\textsf{SNR}_{\text{coop}}$, various bounds are active based on the different configurations and their respective optimization.

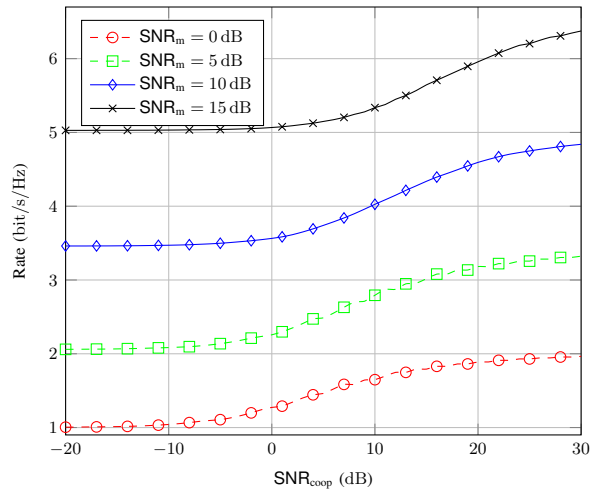
\begin{figure}[t]
	\center
	\begin{tikzpicture}[line width=.75pt,font=\footnotesize,scale=1,every node/.style={scale=0.72}]
		\begin{axis}
			[xmin=-20.0,xmax=30.0,ymin=0.9,ymax=6.7,xlabel={$\textsf{SNR}_{\textrm{coop}}$~($\si{\decibel}$)},ylabel={Rate~(\si{\bit/\second/\hertz})},grid=major,mark options=solid,mark repeat=3,smooth,tension=0.3,legend pos=north west,ylabel style={yshift=-0.4cm},legend cell align=left];
			\addplot[color=red,dashed,mark=o] coordinates{(-20.0,1.005)(-19.0,1.006)(-18.0,1.008)(-17.0,1.009)(-16.0,1.011)(-15.0,1.013)(-14.0,1.016)(-13.0,1.020)(-12.0,1.026)(-11.0,1.033)(-10.0,1.042)(-9.0,1.053)(-8.0,1.067)(-7.0,1.084)(-6.0,1.096)(-5.0,1.107)(-4.0,1.131)(-3.0,1.162)(-2.0,1.199)(-1.0,1.242)(0.0,1.273)(1.0,1.290)(2.0,1.344)(3.0,1.401)(4.0,1.444)(5.0,1.462)(6.0,1.524)(7.0,1.585)(8.0,1.585)(9.0,1.644)(10.0,1.649)(11.0,1.698)(12.0,1.748)(13.0,1.748)(14.0,1.792)(15.0,1.792)(16.0,1.830)(17.0,1.830)(18.0,1.862)(19.0,1.862)(20.0,1.889)(21.0,1.889)(22.0,1.911)(23.0,1.911)(24.0,1.930)(25.0,1.930)(26.0,1.944)(27.0,1.944)(28.0,1.956)(29.0,1.956)(30.0,1.965)};
			\addplot[color=green,dashed,mark=square] coordinates{(-20.0,2.060)(-19.0,2.061)(-18.0,2.062)(-17.0,2.063)(-16.0,2.064)(-15.0,2.066)(-14.0,2.069)(-13.0,2.072)(-12.0,2.076)(-11.0,2.080)(-10.0,2.084)(-9.0,2.090)(-8.0,2.096)(-7.0,2.107)(-6.0,2.120)(-5.0,2.137)(-4.0,2.158)(-3.0,2.183)(-2.0,2.212)(-1.0,2.235)(0.0,2.257)(1.0,2.298)(2.0,2.350)(3.0,2.410)(4.0,2.473)(5.0,2.487)(6.0,2.552)(7.0,2.630)(8.0,2.701)(9.0,2.712)(10.0,2.793)(11.0,2.873)(12.0,2.873)(13.0,2.948)(14.0,2.955)(15.0,3.018)(16.0,3.080)(17.0,3.080)(18.0,3.135)(19.0,3.135)(20.0,3.182)(21.0,3.182)(22.0,3.222)(23.0,3.222)(24.0,3.256)(25.0,3.256)(26.0,3.283)(27.0,3.283)(28.0,3.305)(29.0,3.305)(30.0,3.323)};
			\addplot[color=blue,mark=diamond] coordinates{(-20.0,3.461)(-19.0,3.461)(-18.0,3.461)(-17.0,3.462)(-16.0,3.462)(-15.0,3.463)(-14.0,3.464)(-13.0,3.465)(-12.0,3.467)(-11.0,3.469)(-10.0,3.471)(-9.0,3.474)(-8.0,3.478)(-7.0,3.484)(-6.0,3.490)(-5.0,3.498)(-4.0,3.508)(-3.0,3.520)(-2.0,3.532)(-1.0,3.547)(0.0,3.564)(1.0,3.585)(2.0,3.614)(3.0,3.651)(4.0,3.693)(5.0,3.740)(6.0,3.789)(7.0,3.841)(8.0,3.898)(9.0,3.961)(10.0,4.026)(11.0,4.087)(12.0,4.149)(13.0,4.215)(14.0,4.276)(15.0,4.338)(16.0,4.395)(17.0,4.444)(18.0,4.494)(19.0,4.546)(20.0,4.594)(21.0,4.634)(22.0,4.671)(23.0,4.703)(24.0,4.727)(25.0,4.749)(26.0,4.771)(27.0,4.789)(28.0,4.809)(29.0,4.825)(30.0,4.839)};
			\addplot[color=black,mark=x] coordinates{(-20.0,5.028)(-19.0,5.028)(-18.0,5.028)(-17.0,5.029)(-16.0,5.029)(-15.0,5.029)(-14.0,5.029)(-13.0,5.030)(-12.0,5.030)(-11.0,5.031)(-10.0,5.032)(-9.0,5.033)(-8.0,5.035)(-7.0,5.036)(-6.0,5.038)(-5.0,5.040)(-4.0,5.043)(-3.0,5.047)(-2.0,5.052)(-1.0,5.058)(0.0,5.067)(1.0,5.077)(2.0,5.091)(3.0,5.108)(4.0,5.125)(5.0,5.149)(6.0,5.173)(7.0,5.204)(8.0,5.245)(9.0,5.279)(10.0,5.336)(11.0,5.378)(12.0,5.449)(13.0,5.501)(14.0,5.567)(15.0,5.645)(16.0,5.707)(17.0,5.770)(18.0,5.833)(19.0,5.896)(20.0,5.957)(21.0,6.017)(22.0,6.075)(23.0,6.129)(24.0,6.180)(25.0,6.204)(26.0,6.249)(27.0,6.290)(28.0,6.309)(29.0,6.344)(30.0,6.375)};
			\legend{{$\textsf{SNR}_{\textrm{m}}=\SI{0}{\decibel}$},{$\textsf{SNR}_{\textrm{m}}=\SI{5}{\decibel}$},{$\textsf{SNR}_{\textrm{m}}=\SI{10}{\decibel}$},{$\textsf{SNR}_{\textrm{m}}=\SI{15}{\decibel}$}};
		\end{axis}
	\end{tikzpicture}
	\caption{Comparison of the 3FC scheme for different $\textsf{SNR}_{\textrm{m}}$ values of the symmetric main channel concerning their achievable rate for the Gaussian MC with receiver cooperation.}
	\label{fig:Comparison of the 3FC scheme for different values of the symmetric main channel concerning their achievable rate for the Gaussian MC with receiver cooperation}
\end{figure}

In Fig.~\ref{fig:Comparison of the 3FC scheme for different values of the symmetric main channel concerning their achievable rate for the Gaussian MC with receiver cooperation}, we now suppose that the channel is completely symmetric by adding the condition $\textsf{SNR}_1=\textsf{SNR}_2=\textsf{SNR}_3=\textsf{SNR}_{\textrm{m}}$, meaning that the receivers form a cluster of small size compared to the size of the link from the transmitter to the cluster. The abacus gives some values showing the interest of the proposed scheme, and the gain~\eqref{eq:3FC gain} becomes $G_{\text{3FC}}^\textrm{m}=\log\left(1+\frac{2\textsf{SNR}_{\textrm{m}}}{1+\textsf{SNR}_{\textrm{m}}}\right)$ and grows to $\log(3)$ as $\textsf{SNR}_{\textrm{m}}\rightarrow\infty$.

We have underlined a number of rules that are, 1)~use superpositions of CFs and DFs to obtain a low latency architecture for the scheme, 2)~obtain bounds with a good structure in mutual information for the two extreme cases, i.e., when the cooperation link is weak and strong by using DFs and CFs respectively, and by letting information flow properly through each node, 3)~DFs can only be used when short messages are used, and refine information about a short message after that the decoding step has already been performed is of no use, thus CF should be used before DF on a given short message at a given node, 4)~perform the CFs in the first round in an ``all to one structure'' and do not use it further on short messages, and 5)~approach the probability distribution of the cutset upper bound by using DF in a ``chain structure'' to exploit the correlation of the codebooks between the receivers and the transmitter.

\section{Summary and Discussion}
\label{sec:Summary and Discussion}

In this paper, we investigated the impact of receiver cooperation on the throughput of a three-receiver MC. We proposed a fully interactive cooperation scheme based on an information-theoretic analysis that remains tractable. We showed through numerical results focusing on the SISO Gaussian MC that our proposed 3FC scheme outperforms existing schemes in which no interaction is exploited or in which information does not flow properly through each node. This asymmetric interaction comes from the specific superpositions of CF and DF at the transmitter and receivers that we developed, and permits to enlarge the achievable rate while preserving a low latency. The CF forms a ``all to one structure'' to initiate the scheme, and the DF forms a ``chain structure'' to allow a correlation of each layer of the superposition in cooperation with the transmitter toward the next receiver in the chain. Our results revealed that interaction is particularly helpful in comparison to the NNC and the 3PC when the main channel has a slight asymmetry. When the main channel is symmetric the 3FC is equal to the NNC, while when the main channel is very asymmetric the 3FC tends to the 3PC.

\begin{figure}[t]
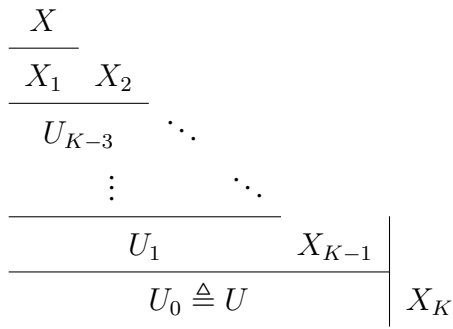

	\centering
	\begin{tabular}{ccccc|c}
		$X$&\\\cline{1-1}
		$X_1$&$X_2$\\\cline{1-2}
		\multicolumn{2}{c}{$U_{K-3}$}&$\ddots$\\
		\multicolumn{3}{c}{$\vdots$}&$\ddots$\\\cline{1-4}
		\multicolumn{4}{c}{$U_1$}&$X_{K-1}$\\\cline{1-5}
		\multicolumn{5}{c|}{$U_0\triangleq U$}&$X_K$
	\end{tabular}
	\caption{Superposition structure for $K$~receivers.}
	\label{fig:Superposition structure for K receivers}
\end{figure}

The bounds in the general case of $K\geq 2$~receivers eludes us because of 1)~the complexity induced by the sliding windows of increasing size that have to be handled for each new receiver performing DF that is added to the system, and 2)~the chain rules and the Fourier-Motzkin elimination procedure that would have to be applied on the bounds to get a closed-form expression. Even with the closed-form expression, it is doubtful that an easy comparison would be possible between the expression of the $K$-receiver fully interactive cooperation scheme~($K$FC) and, e.g., the NNC, due to the inherent differences of the bounds and of their respective probability distribution, and to the complexity of the numerical comparison since the number of parameters to optimize would quickly increase. The gain of such a scheme as defined in~\eqref{eq:3FC gain} would be $G_{K\text{FC}}=\log\left(1+\frac{\sum_{k=1}^{K-1}\textsf{SNR}_k}{1+\textsf{SNR}_K}\right)$, and in the completely symmetric case, $G_{K\text{FC}}^\textrm{m}$ would grow to $\log(K)$ as $\textsf{SNR}_{\textrm{m}}\rightarrow\infty$. However, as a result of our work, we can give the structure of the $K$FC. In the $K$FC, the transmitter multicasts a short message and then sequentially, 1)~in the first round, all the receivers except the strongest use CF toward the strongest receiver, labeled receiver~1, and 2)~recursively, in each of the following $K-1$~rounds, e.g., round $r\in[2:K]$, all the $r-1$ receivers that have already decoded the current message cooperate with the transmitter (use a correlated layer in their stack of superposition) by using DF toward receiver~$r$. Note that all the receivers using DF have to perform a sliding window decoding, and that receiver~$K$ can perform either a sliding window decoding (in this case the latency is only of $K$~blocks) or a backward decoding since it does not use DF. It follows that the probability distribution is
\begin{align}
	p(u_0)\prod_{k=1}^{K-3}p(u_k\vert u_{k-1})p(x_1\vert u_{K-3})p(x\vert x_1)\prod_{k=2}^K p(x_k\vert u_{K-1-k})\prod_{k=2}^K p(\tilde{y}_k\vert x_k,y_k),\label{eq:general probability distribution}
\end{align}
where $\prod_{K=a}^b$ exists only if $a\leq b$ and $u_0\triangleq u$. The superposition structure defined by \eqref{eq:general probability distribution} is illustrated in Fig.~\ref{fig:Superposition structure for K receivers}. The transmitter multicasts codewords with the structure presented in the first column. The receiver~1 generates codewords with the structure presented in the same column but without the last upper layer since it has already decoded all the previous messages but not the current one. Recursively for all the remaining receivers, every receiver has one less DF layer until finally the last receiver is reached. The last receiver does not have a DF layer since it does not perform DF toward any other receiver. Each DF layer is identical for all the receivers that have already decoded the corresponding message since they share the same information, and thus can construct the same clouds in the codeword and correlate it accordingly. All the receivers except the strongest present a CF layer. Those layers may differ from one receiver to the other, since no decoding operation has been performed on this information yet. Those layers are intended to all the receivers that are stronger than the ones transmitting it, and will be stored and used in later rounds to decode the corresponding messages.

More recently, the distributed decode-forward~(DDF)~\cite{inproceedings11} and even more the generalization of the NNC and DDF called the NNC with partial DF~(NNC-PDF)~\cite{unpublished9} have been proposed for similar networks. Such schemes seem promising since they exploit synergies between PDFs and PCDFs, respectively. Unfortunately, the achievable rate regions in such schemes involve some auxiliary random variable which makes their evaluation and fair comparison to our results extremely complicated. Nevertheless, it remains an interesting future direction of investigation.

\appendices

\section{Description of the Proposed Scheme}
\label{appx:Description of the Proposed Scheme}

\begin{figure*}[t]
	\centering
	\begin{tikzpicture}[line width=.75pt,font=\footnotesize]
		\coordinate(Messagetx)at(0,0);
		\coordinate(Encoder)at(1,0);
		\coordinate(Encoderbl)at([shift={(0,-.3)}]Encoder);
		\coordinate(Encoderar)at([shift={(1.5,.3)}]Encoder);
		\coordinate(Encoderr)at([shift={(1.5,0)}]Encoder);
		\coordinate(Channell)at(6,0);
		\coordinate(Channelr)at(8,0);
		\coordinate(Channelarout)at([shift={(0,1.65)}]Channelr);
		\coordinate(Channelarin)at([shift={(0,1)}]Channelr);
		\coordinate(Channelmrout)at([shift={(0,.35)}]Channelr);
		\coordinate(Channelmrin)at([shift={(0,-.3)}]Channelr);
		\coordinate(Channelbrout)at([shift={(0,-.95)}]Channelr);
		\coordinate(Channelbrin)at([shift={(0,-1.6)}]Channelr);
		
		\coordinate(Decoder1)at(11,1.65);
		\coordinate(Decoder1bl)at([shift={(-1.5,-.3)}]Decoder1);
		\coordinate(Decoder1ar)at([shift={(0,.3)}]Decoder1);
		\coordinate(Decoder1l)at([shift={(-1.5,0)}]Decoder1);
		\coordinate(Decoder1coop)at([shift={(-.75,-.3)}]Decoder1);
		\coordinate(Message1rx)at(12,1.65);
		
		\coordinate(Decoder2)at(11,.35);
		\coordinate(Decoder2bl)at([shift={(-1.5,-.3)}]Decoder2);
		\coordinate(Decoder2ar)at([shift={(0,.3)}]Decoder2);
		\coordinate(Decoder2l)at([shift={(-1.5,0)}]Decoder2);
		\coordinate(Decoder2coop)at([shift={(-.75,-.3)}]Decoder2);
		\coordinate(Message2rx)at(12,.35);
		
		\coordinate(Decoder3)at(11,-.95);
		\coordinate(Decoder3bl)at([shift={(-1.5,-.3)}]Decoder3);
		\coordinate(Decoder3ar)at([shift={(0,.3)}]Decoder3);
		\coordinate(Decoder3l)at([shift={(-1.5,0)}]Decoder3);
		\coordinate(Decoder3coop)at([shift={(-.75,-.3)}]Decoder3);
		\coordinate(Message3rx)at(12,-.95);
		
		\coordinate(Channelbl)at([shift={(0,-1.95)}]Channell);
		\coordinate(Channelar)at([shift={(0,1.95)}]Channelr);
		\draw[->](Messagetx)node[left]{\begin{tabular}{@{}c@{}}$M_j,$\\$j\in[1:b-3]$\end{tabular}}--(Encoder);
		\draw(Encoderbl)rectangle(Encoderar)node[pos=.5]{TX};
		\draw[->](Encoderr)--(Channell)node[above,midway]{$X^n(m_j\vert m_{j-2}\vert m_{j-3})$};
		\draw[->](Channelarout)--(Decoder1l)node[above,midway]{$Y_1^n(j)$};
		\draw(Decoder1bl)rectangle(Decoder1ar)node[pos=.5]{RX~1};
		\draw[->](Decoder1coop)--(Channelarin)node[yshift=-5,right,near start]{$X_1^n(\hat{m}_{j-2,1}\vert\hat{m}_{j-3,1})$};
		\draw[->](Decoder1)--(Message1rx)node[right]{\begin{tabular}{@{}c@{}}$\hat{M}_{j-1,1},$\\$j\in[2:b-2]$\end{tabular}};
		
		\draw[->](Channelmrout)--(Decoder2l)node[above,midway]{$Y_2^n(j)$};
		\draw(Decoder2bl)rectangle(Decoder2ar)node[pos=.5]{RX~2};
		\draw[->](Decoder2coop)--(Channelmrin)node[yshift=-5,right,near start]{$X_2^n(l_{j-1,2}\vert\hat{m}_{j-3,2})$};
		\draw[->](Decoder2)--(Message2rx)node[right]{\begin{tabular}{@{}c@{}}$\hat{M}_{j-2,2},$\\$j\in[3:b-1]$\end{tabular}};
		
		\draw[->](Channelbrout)--(Decoder3l)node[above,midway]{$Y_3^n(j)$};
		\draw(Decoder3bl)rectangle(Decoder3ar)node[pos=.5]{RX~3};
		\draw[->](Decoder3coop)--(Channelbrin)node[yshift=-5,right,near start]{$X_3^n(l_{j-1,3})$};
		\draw[->](Decoder3)--(Message3rx)node[yshift=-6,right]{\begin{tabular}{@{}c@{}}$\hat{M}_{j-3,3},$\\$j\in[4:b],$\\if $\hat{M}_{j-2,3},\hat{M}_{j,3}$\end{tabular}};
		
		\draw(Channelbl)rectangle(Channelar)node[pos=.5]{\begin{tabular}{@{}c@{}}$P_{Y_1,Y_2,Y_3\vert}$\\$_{X,X_1,X_2,X_3}$\end{tabular}};
	\end{tikzpicture}
	\caption{The MC with receiver cooperation for the 3FC scheme $\textrm{STG}_{1,2,3}^{(2,3)}$.}
	\label{fig:The MC with receiver cooperation for the 3FC scheme}
\end{figure*}
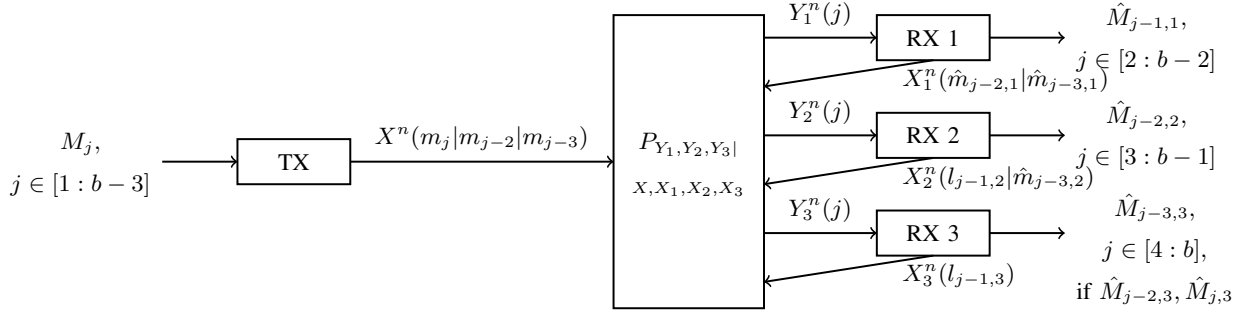

\begin{table*}[t]
	\centering
	\begin{tabular}{c|cccc}
		Block&$j$&$j+1$&$j+2$&$j+3$\\
		\hline\hline
		$U$&$u^n(m_{j-3})$&$u^n(m_{j-2})$&$u^n(m_{j-1})$&$u^n(\tikzmark{j+3-src-cloud}\underline{m_j})$\\
		$X$&$x^n(\tikzmark{j-src}\underline{m_j}\vert m_{j-2}\vert m_{j-3})$&$x^n(m_{j+1}\vert m_{j-1}\vert m_{j-2})$&$x^n(m_{j+2}\vert\tikzmark{j+2-src}\underline{m_j}\vert m_{j-1})$&$x^n(m_{j+3}\vert m_{j+1}\vert\tikzmark{j+3-src}\underline{m_j})$\\
		\hline
		$Y_1$&$\tikzmark{j-dst1}\underline{y_1^n(j)}$&$\tikzmark{j+1-dst1}\underline{y_1^n(j+1)}$&$y_1^n(j+2)$&$y_1^n(j+3)$\\
		$X_1$&$x_1^n(\hat{m}_{j-2,1}\vert\hat{m}_{j-3,1})$&$x_1^n(\hat{m}_{j-1,1}\vert\hat{m}_{j-2,1})$&$x_1^n(\tikzmark{j+2-src1}\underline{\hat{m}_{j,1}}\vert\hat{m}_{j-1,1})$&$x_1^n(\hat{m}_{j+1,1}\vert\tikzmark{j+3-src1}\underline{\hat{m}_{j,1}})$\\
		$\hat{Y}_1$&\begin{tabular}{@{}c@{}}$\hat{m}_{j-1,1},\hat{l}_{j-1,2},$\\$\hat{k}_{j-1,2},\hat{l}_{j-1,3},\hat{k}_{j-1,3}$\end{tabular}&\begin{tabular}{@{}c@{}}$\tikzmark{j+1-dec1}\underline{\hat{m}_{j,1}},\underline{\hat{l}_{j,2}},$\\$\underline{\hat{k}_{j,2}},\underline{\hat{l}_{j,3}},\underline{\hat{k}_{j,3}}$\end{tabular}&\begin{tabular}{@{}c@{}}$\hat{m}_{j+1,1},\hat{l}_{j+1,2},$\\$\hat{k}_{j+1,2},\hat{l}_{j+1,3},\hat{k}_{j+1,3}$\end{tabular}&\begin{tabular}{@{}c@{}}$\hat{m}_{j+2,1},\hat{l}_{j+2,2},$\\$\hat{k}_{j+2,2},\hat{l}_{j+2,3},\hat{k}_{j+2,3}$\end{tabular}\\
		\hline
		$Y_2$&$\tikzmark{j-dst2}\underline{y_2^n(j)}$&$\tikzmark{j+1-dst2}\underline{y_2^n(j+1)}$&$\tikzmark{j+2-dst2}\underline{y_2^n(j+2)}$&$y_2^n(j+3)$\\
		$\tilde{Y}_2$&\begin{tabular}{@{}c@{}}$\tilde{y}_2^n(\tikzmark{j-dst2.1}\underline{k_{j,2}}\vert l_{j-1,2}\vert\hat{m}_{j-3,2}),$\\$\tikzmark{j-dst2.2}\underline{l_{j,2}}$\end{tabular}&\begin{tabular}{@{}c@{}}$\tilde{y}_2^n(k_{j+1,2}\vert l_{j,2}\vert\hat{m}_{j-2,2}),$\\$l_{j+1,2}$\end{tabular}&\begin{tabular}{@{}c@{}}$\tilde{y}_2^n(k_{j+2,2}\vert l_{j+1,2}\vert\hat{m}_{j-1,2}),$\\$l_{j+2,2}$\end{tabular}&\begin{tabular}{@{}c@{}}$\tilde{y}_2^n(k_{j+3,2}\vert l_{j+2,2}\vert\hat{m}_{j,2}),$\\$l_{j+3,2}$\end{tabular}\\
		$X_2$&$x_2^n(l_{j-1,2}\vert\hat{m}_{j-3,2})$&$x_2^n(\tikzmark{j+1-src2}\underline{l_{j,2}}\vert\hat{m}_{j-2,2})$&$x_2^n(l_{j+1,2}\vert\hat{m}_{j-1,2})$&$x_2^n(l_{j+2,2}\vert\tikzmark{j+3-src2}\underline{\hat{m}_{j,2}})$\\
		$\hat{Y}_2$&$\hat{m}_{j-2,2},\hat{l}_{j-1,3},\hat{k}_{j-1,3}$&$\hat{m}_{j-1,2},\underline{\hat{l}_{j,3}},\underline{\hat{k}_{j,3}}$&$\tikzmark{j+2-dec2}\underline{\hat{m}_{j,2}},\hat{l}_{j+1,3},\hat{k}_{j+1,3}$&$\hat{m}_{j+1,2},\hat{l}_{j+2,3},\hat{k}_{j+2,3}$\\
		\hline
		$Y_3$&$\tikzmark{j-dst3}\underline{y_3^n(j)}$&$y_3^n(j+1)$&$y_3^n(j+2)$&$\tikzmark{j+3-dst3}\underline{y_3^n(j+3)}$\\
		$\tilde{Y}_3$&\begin{tabular}{@{}c@{}}$\tilde{y}_3^n(\tikzmark{j-dst3.1}\underline{k_{j,3}}\vert l_{j-1,3}),$\\$\tikzmark{j-dst3.2}\underline{l_{j,3}}$\end{tabular}&\begin{tabular}{@{}c@{}}$\tilde{y}_3^n(k_{j+1,3}\vert l_{j,3}),$\\$l_{j+1,3}$\end{tabular}&\begin{tabular}{@{}c@{}}$\tilde{y}_3^n(k_{j+2,3}\vert l_{j+1,3}),$\\$l_{j+2,3}$\end{tabular}&\begin{tabular}{@{}c@{}}$\tilde{y}_3^n(k_{j+3,3}\vert l_{j+2,3}),$\\$l_{j+3,3}$\end{tabular}\\
		$X_3$&$x_3^n(l_{j-1,3})$&$x_3^n(\tikzmark{j+1-src3}\underline{l_{j,3}})$&$x_3^n(l_{j+1,3})$&$x_3^n(l_{j+2,3})$\\
		$\hat{Y}_3$&\begin{tabular}{@{}c@{}}$\hat{m}_{j-3,3},$\\if $\hat{m}_{j-2,3},\hat{m}_{j,3}$\end{tabular}&\begin{tabular}{@{}c@{}}$\hat{m}_{j-2,3},$\\if $\hat{m}_{j-1,3},\hat{m}_{j+1,3}$\end{tabular}&\begin{tabular}{@{}c@{}}$\hat{m}_{j-1,3},$\\if $\hat{m}_{j,3},\hat{m}_{j+2,3}$\end{tabular}&\begin{tabular}{@{}c@{}}$\tikzmark{j+3-dec3}\underline{\hat{m}_{j,3}},$\\if $\hat{m}_{j+1,3},\hat{m}_{j+3,3}$\end{tabular}\\
		\hline
	\end{tabular}
	\begin{tikzpicture}[overlay,remember picture,yshift=.25pt,shorten >=.5pt,shorten <=.5pt,stripeal/.style={draw, pattern=north east lines},stripear/.style={draw, pattern=north west lines},stripea/.style={draw, pattern=horizontal lines},stripel/.style={draw, pattern=vertical lines}]
		\tikzstyle{DashedThick}=[dashed,thick];
		\draw[->]([xshift=4pt,yshift=-2pt]{pic cs:j-src})[bend right]to({pic cs:j-dst1});
		\draw[->]([xshift=4pt,yshift=-2pt]{pic cs:j-src})[bend right]to({pic cs:j-dst2});
		\draw[->]([xshift=4pt,yshift=-2pt]{pic cs:j-src})[bend right]to({pic cs:j-dst3});
		\draw[stripeal,opacity=0.3]([xshift=0pt,yshift=-4pt]{pic cs:j-dst2.1})rectangle([xshift=15pt,yshift=7pt]{pic cs:j-dst2.1});
		\draw[stripeal,opacity=0.3]([xshift=0pt,yshift=-4pt]{pic cs:j-dst2.2})rectangle([xshift=13pt,yshift=7pt]{pic cs:j-dst2.2});
		\draw[stripeal,opacity=0.3]([xshift=0pt,yshift=-4pt]{pic cs:j+1-src2})rectangle([xshift=13pt,yshift=7pt]{pic cs:j+1-src2});
		\draw[stripeal,opacity=0.3]([xshift=0pt,yshift=-4pt]{pic cs:j+1-dst1})rectangle([xshift=38pt,yshift=7pt]{pic cs:j+1-dst1});
		\draw[stripear,opacity=0.3]([xshift=0pt,yshift=-4pt]{pic cs:j-dst3.1})rectangle([xshift=15pt,yshift=7pt]{pic cs:j-dst3.1});
		\draw[stripear,opacity=0.3]([xshift=0pt,yshift=-4pt]{pic cs:j-dst3.2})rectangle([xshift=13pt,yshift=7pt]{pic cs:j-dst3.2});
		\draw[stripear,opacity=0.3]([xshift=0pt,yshift=-4pt]{pic cs:j+1-src3})rectangle([xshift=13pt,yshift=7pt]{pic cs:j+1-src3});
		\draw[stripear,opacity=0.3]([xshift=0pt,yshift=-4pt]{pic cs:j+1-dst1})rectangle([xshift=38pt,yshift=7pt]{pic cs:j+1-dst1});
		\draw[stripear,opacity=0.3]([xshift=0pt,yshift=-4pt]{pic cs:j+1-dst2})rectangle([xshift=38pt,yshift=7pt]{pic cs:j+1-dst2});
		\draw[stripea,opacity=0.2]([xshift=0pt,yshift=-4pt]{pic cs:j+2-src})rectangle([xshift=13pt,yshift=7pt]{pic cs:j+2-src});
		\draw[stripea,opacity=0.2]([xshift=0pt,yshift=-4pt]{pic cs:j+2-dst2})rectangle([xshift=38pt,yshift=7pt]{pic cs:j+2-dst2});
		\draw[stripea,opacity=0.2]([xshift=0pt,yshift=-4pt]{pic cs:j+1-dec1})rectangle([xshift=18pt,yshift=7pt]{pic cs:j+1-dec1});
		\draw[stripea,opacity=0.2]([xshift=0pt,yshift=-4pt]{pic cs:j+2-src1})rectangle([xshift=18pt,yshift=7pt]{pic cs:j+2-src1});
		\draw[->,thick]([xshift=0pt,yshift=0pt]{pic cs:j+2-dst2})--([xshift=-32pt,yshift=0pt]{pic cs:j+2-dst2})--([xshift=-15pt,yshift=0pt]{pic cs:j+2-dec2})--([xshift=0pt,yshift=0pt]{pic cs:j+2-dec2});
		\draw[->,thick]([xshift=0pt,yshift=0pt]{pic cs:j+1-dst1})--([xshift=-21pt,yshift=0pt]{pic cs:j+1-dst1})--([xshift=-21pt,yshift=0pt]{pic cs:j+1-dec1})--([xshift=0pt,yshift=0pt]{pic cs:j+1-dec1});
		\draw[stripel,opacity=0.2]([xshift=0pt,yshift=-4pt]{pic cs:j+3-src-cloud})rectangle([xshift=13pt,yshift=7pt]{pic cs:j+3-src-cloud});
		\draw[stripel,opacity=0.2]([xshift=0pt,yshift=-4pt]{pic cs:j+3-src})rectangle([xshift=13pt,yshift=7pt]{pic cs:j+3-src});
		\draw[stripel,opacity=0.2]([xshift=0pt,yshift=-4pt]{pic cs:j+3-dst3})rectangle([xshift=38pt,yshift=7pt]{pic cs:j+3-dst3});
		\draw[stripel,opacity=0.2]([xshift=0pt,yshift=-4pt]{pic cs:j+1-dec1})rectangle([xshift=18pt,yshift=7pt]{pic cs:j+1-dec1});
		\draw[stripel,opacity=0.2]([xshift=0pt,yshift=-4pt]{pic cs:j+3-src1})rectangle([xshift=18pt,yshift=7pt]{pic cs:j+3-src1});
		\draw[stripel,opacity=0.2]([xshift=0pt,yshift=-4pt]{pic cs:j+2-dec2})rectangle([xshift=18pt,yshift=7pt]{pic cs:j+2-dec2});
		\draw[stripel,opacity=0.2]([xshift=0pt,yshift=-4pt]{pic cs:j+3-src2})rectangle([xshift=18pt,yshift=7pt]{pic cs:j+3-src2});
		\draw[->,thick]([xshift=0pt,yshift=0pt]{pic cs:j+3-dst3})--([xshift=-17pt,yshift=0pt]{pic cs:j+3-dst3})--([xshift=-26pt,yshift=0pt]{pic cs:j+3-dec3})--([xshift=0pt,yshift=0pt]{pic cs:j+3-dec3});
  \end{tikzpicture}
	\caption{Encoding, transmission, quantization distortion, and decoding for the MC with receiver cooperation for the 3FC scheme $\textrm{STG}_{1,2,3}^{(2,3)}$. The table focuses on the message $\underline{m_j}$ and its representations. The curved arrows correspond to the first multicast of the message $\underline{m_j}$. The oblique stripes represent CF operations while vertical and horizontal ones represent DF operations. The thick arrows correspond to the decoding steps.}
	\label{tab:Encoding, transmission, quantization distortion, and decoding for the MC with receiver cooperation for the 3FC scheme}
\end{table*}

We consider $\textrm{STG}_{1,2,3}^{(2,3)}$ illustrated in Fig.~\ref{fig:The MC with receiver cooperation for the 3FC scheme}. In Tab.~\ref{tab:Encoding, transmission, quantization distortion, and decoding for the MC with receiver cooperation for the 3FC scheme}, the encoding and decoding related to the short message $m_j$ are underlined, and the thick arrows correspond to the decoding steps. The patterns in Tab.~\ref{tab:Encoding, transmission, quantization distortion, and decoding for the MC with receiver cooperation for the 3FC scheme} and Fig.~\ref{fig:The MC with receiver cooperation for the 3FC and 3PC schemes - 3FC representation} represent, 1)~CF operations for the oblique stripes at respectively receiver~2 (\begin{tikzpicture}[stripeal/.style={draw, pattern=north east lines},stripear/.style={draw, pattern=north west lines},stripea/.style={draw, pattern=horizontal lines},stripel/.style={draw, pattern=vertical lines}]
	\coordinate(belowbelowleft)at(0,0);
	\coordinate(aboveright)at(0.3,0.3);
	\draw[stripeal,opacity=0.3](belowbelowleft)rectangle(aboveright);
\end{tikzpicture}) and~3 (\begin{tikzpicture}[stripeal/.style={draw, pattern=north east lines},stripear/.style={draw, pattern=north west lines},stripea/.style={draw, pattern=horizontal lines},stripel/.style={draw, pattern=vertical lines}]
	\coordinate(belowbelowleft)at(0,0);
	\coordinate(aboveright)at(0.3,0.3);
	\draw[stripear,opacity=0.3](belowbelowleft)rectangle(aboveright);
\end{tikzpicture}), and 2)~DF operations for the vertical and horizontal stripes toward receiver~2 (\begin{tikzpicture}[stripeal/.style={draw, pattern=north east lines},stripear/.style={draw, pattern=north west lines},stripea/.style={draw, pattern=horizontal lines},stripel/.style={draw, pattern=vertical lines}]
	\coordinate(belowbelowleft)at(0,0);
	\coordinate(aboveright)at(0.3,0.3);
	\draw[stripea,opacity=0.2](belowbelowleft)rectangle(aboveright);
\end{tikzpicture}) and~3 (\begin{tikzpicture}[stripeal/.style={draw, pattern=north east lines},stripear/.style={draw, pattern=north west lines},stripea/.style={draw, pattern=horizontal lines},stripel/.style={draw, pattern=vertical lines}]
	\coordinate(belowbelowleft)at(0,0);
	\coordinate(aboveright)at(0.3,0.3);
	\draw[stripel,opacity=0.2](belowbelowleft)rectangle(aboveright);
\end{tikzpicture}), with the same respective patterns. A sequence of $(b-3)$~messages $M_j,\ j\in[1:b-3]$, are selected independently and uniformly over $[1:2^{nR}]$ and are separately encoded and transmitted over $b$~blocks. The average rate $R\frac{b-3}{b}$ tends to $R$ as $b\rightarrow\infty$. The set of weak $\epsilon$-typical $n$-sequences $\mathcal{T}_{\epsilon}^{(n)}$ used are defined as in~\cite{book1,book3}; this permits to apply continuous probability distributions to our bounds.

\emph{Codebook generation.} Fix the probability distribution,
\begin{align}
	p(u)p(x_1\vert u)p(x\vert x_1)p(x_2\vert u)p(x_3)p(y_1,y_2,y_3\vert x,x_1,x_2,x_3)p(\tilde{y}_2\vert x_2,y_2)p(\tilde{y}_3\vert x_3,y_3).
\end{align}
Generate at random an independent codebook for each block (only four such independent codebooks used for every consecutive quadruple-block are required, so that joint decoding over any four adjacent blocks result in independent error events). For $j\in[1:b]$, randomly and independently generate $2^{nR}$ sequences $u^n(m_{j-3}),\ m_{j-3}\in[1:2^{nR}]$, each according to $\prod_{i=1}^n p_{U}(u_i)$. For each $m_{j-3}\in[1:2^{nR}]$, randomly and conditionally independently generate $2^{nR}$ sequences $x_1^n(m_{j-2}\vert m_{j-3}),\ m_{j-2}\in[1:2^{nR}]$, each according to $\prod_{i=1}^n p_{X_1\vert U}(x_{1i}\vert u_i(m_{j-3}))$. For each $(m_{j-2},m_{j-3})\in[1:2^{nR}]^2$, randomly and conditionally independently generate $2^{nR}$ sequences $x^n(m_j\vert m_{j-2}\vert m_{j-3}),\ m_j\in[1:2^{nR}]$, each according to $\prod_{i=1}^n p_{X\vert X_1}(x_i\vert x_{1i}(m_{j-2}\vert m_{j-3}))$. For each $m_{j-3}\in[1:2^{nR}]$, randomly and conditionally independently generate $2^{nR_2}$ sequences $x_2^n(l_{j-1,2}\vert m_{j-3}),\ l_{j-1,2}\in[1:2^{nR_2}]$, each according to $\prod_{i=1}^n p_{X_2\vert U}(x_{2i}\vert u_i(m_{j-3}))$. For each $(l_{j-1,2},m_{j-3})\in[1:2^{nR_2}]\times[1:2^{nR}]$, randomly and conditionally independently generate $2^{n\tilde{R}_2}$ sequences $\tilde{y}_2^n(k_{j,2}\vert l_{j-1,2}\vert m_{j-3}),\ k_{j,2}\in[1:2^{n\tilde{R}_2}]$, each according to $\prod_{i=1}^n p_{\tilde{Y}_2\vert X_2}(\tilde{y}_{2i}\vert x_{2i}(l_{j-1,2}\vert m_{j-3}))$. Randomly and independently generate $2^{nR_3}$ sequences $x_3^n(l_{j-1,3}),\ l_{j-1,3}\in[1:2^{nR_3}]$, each according to $\prod_{i=1}^n p_{X_3}(x_{3i})$. For each $l_{j-1,3}\in[1:2^{nR_3}]$, randomly and conditionally independently generate $2^{n\tilde{R}_3}$ sequences $\tilde{y}_3^n(k_{j,3}\vert l_{j-1,3}),\ k_{j,3}\in[1:2^{n\tilde{R}_3}]$, each according to $\prod_{i=1}^n p_{\tilde{Y}_3\vert X_3}(\tilde{y}_{3i}\vert x_{3i}(l_{j-1,3}))$.

The codebooks are defined as,
\begin{multline}
	\mathcal{C}_j=\Big\{(u^n(m_{j-3}),x^n(m_j\vert m_{j-2}\vert m_{j-3}),x_1^n(m_{j-2}\vert m_{j-3}),x_2^n(l_{j-1,2}\vert m_{j-3}),\tilde{y}_2^n(k_{j,2}\vert l_{j-1,2}\vert m_{j-3}),\\
	x_3^n(l_{j-1,3}),\tilde{y}_3^n(k_{j,3}\vert l_{j-1,3}))\vert m_j,m_{j-2},m_{j-3}\in[1:2^{nR}],\\
	l_{j-1,2}\in[1:2^{nR_2}],k_{j,2}\in[1:2^{n\tilde{R}_2}],l_{j-1,3}\in[1:2^{nR_3}],k_{j,3}\in[1:2^{n\tilde{R}_3}]\Big\},
\end{multline}
for $j\in[1:b]$. Partition the set $[1:2^{n\tilde{R}_2}]$ into $2^{nR_2}$ equal size bins $\mathcal{B}(l_{j,2})=[(l_{j,2}-1)2^{n(\tilde{R}_2-R_2)}+1:l_{j,2}2^{n(\tilde{R}_2-R_2)}],\ l_{j,2}\in[1:2^{nR_2}],\ \tilde{R}_2\geq R_2$. Partition the set $[1:2^{n\tilde{R}_3}]$ into $2^{nR_3}$ equal size bins $\mathcal{B}(l_{j,3})=[(l_{j,3}-1)2^{n(\tilde{R}_3-R_3)}+1:l_{j,3}2^{n(\tilde{R}_3-R_3)}],\ l_{j,3}\in[1:2^{nR_3}],\ \tilde{R}_3\geq R_3$. The codebooks and the bin assignments are revealed to all parties.

\emph{Encoding.} Let $m_j\in[1:2^{nR}]$ be the message to be sent over the block~$j$. The encoder transmits $x^n(m_j\vert m_{j-2}\vert m_{j-3})$ from the codebook $\mathcal{C}_j$, of first cloud center $u^n(m_{j-3})$, where $m_{-2}=m_{-1}=m_0=m_{b-2}=m_{b-1}=m_b=1$ by convention.

\emph{Relay encoding at receiver~2.} Let $l_{0,2}=l_{b-2,2}=l_{b-1,2}=1$ and $\hat{m}_{-2,2}=\hat{m}_{-1,2}=\hat{m}_{0,2}=1$ by convention. At the end of block~$j$, the relay receiver~2 finds an index $k_{j,2}$ s.t.,
\begin{align}
	(y_2^n(j),\tilde{y}_2^n(k_{j,2}\vert l_{j-1,2}\vert\hat{m}_{j-3,2}),x_2^n(l_{j-1,2}\vert\hat{m}_{j-3,2}),u^n(\hat{m}_{j-3,2}))\in\mathcal{T}_{\epsilon'}^{(n)}.
\end{align}
If there is more than one such index, it selects one of them uniformly at random. If there is no such index, it selects an index from $[1:2^{n\tilde{R}_2}]$ uniformly at random. In block~$j+1$ the relay receiver~2 transmits $x_2^n(l_{j,2}\vert\hat{m}_{j-2,2})$ from codebook $\mathcal{C}_{j+1}$, where $k_{j,2}\in\mathcal{B}(l_{j,2})$, and $\hat{m}_{j-2,2}$ was decoded in block~$j$.

\emph{Relay encoding at receiver~3.} Let $l_{0,3}=l_{b-2,3}=l_{b-1,3}=1$ by convention. At the end of block~$j$, the relay receiver~3 finds an index $k_{j,3}$ s.t.\ $(y_3^n(j),\tilde{y}_3^n(k_{j,3}\vert l_{j-1,3}),x_3^n(l_{j-1,3}))\in\mathcal{T}_{\epsilon'}^{(n)}$. If there is more than one such index, it selects one of them uniformly at random. If there is no such index, it selects an index from $[1:2^{n\tilde{R}_3}]$ uniformly at random. In block~$j+1$ the relay receiver~3 transmits $x_3^n(l_{j,3})$ from codebook $\mathcal{C}_{j+1}$, where $k_{j,3}\in\mathcal{B}(l_{j,3})$.

\emph{Decoding at receiver~1.} Let $\epsilon>\epsilon'$. At the end of block~$j+1$, the decoder receiver~1 finds the unique pair of indices $(\hat{l}_{j,2},\hat{l}_{j,3})$ s.t.,
\begin{align}
	(x_2^n(\hat{l}_{j,2}\vert\hat{m}_{j-2,1}),x_3^n(\hat{l}_{j,3}),x_1^n(\hat{m}_{j-1,1}\vert\hat{m}_{j-2,1}),y_1^n(j+1),u^n(\hat{m}_{j-2,1}))\in\mathcal{T}_{\epsilon'}^{(n)}.
\end{align}
It then finds the unique message $\hat{m}_{j,1}$ s.t.,
\begin{multline}
	(x^n(\hat{m}_{j,1}\vert\hat{m}_{j-2,1}\vert\hat{m}_{j-3,1}),x_2^n(\hat{l}_{j-1,2}\vert\hat{m}_{j-3,1}),\tilde{y}_2^n(\hat{k}_{j,2}\vert\hat{l}_{j-1,2}\vert\hat{m}_{j-3,1}),\\
	x_3^n(\hat{l}_{j-1,3}),\tilde{y}_3^n(\hat{k}_{j,3}\vert\hat{l}_{j-1,3}),x_1^n(\hat{m}_{j-2,1}\vert\hat{m}_{j-3,1}),y_1^n(j),u^n(\hat{m}_{j-3,1}))\in\mathcal{T}_{\epsilon}^{(n)},
\end{multline}
for some $\hat{k}_{j,2}\in\mathcal{B}(\hat{l}_{j,2})$ and $\hat{k}_{j,3}\in\mathcal{B}(\hat{l}_{j,3})$.

\emph{Relay encoding at receiver~1.} Let $\hat{m}_{-2,1}=\hat{m}_{-1,1}=\hat{m}_{0,1}=\hat{m}_{b-2,1}=1$ by convention. In block~$j+2$ the relay receiver~1 transmits $x_1^n(\hat{m}_{j,1}\vert\hat{m}_{j-1,1})$ from the codebook $\mathcal{C}_{j+2}$.

\emph{Sliding window decoding at receiver~2.} Let $\epsilon>\epsilon'$. At the end of block~$j+1$, the decoder receiver~2 finds the unique index $\hat{l}_{j,3}$ s.t.,
\begin{align}
	(x_3^n(\hat{l}_{j,3}),x_1^n(\hat{m}_{j-1,2}\vert\hat{m}_{j-2,2}),x_2^n(l_{j,2}\vert\hat{m}_{j-2,2}),y_2^n(j+1),u^n(\hat{m}_{j-2,2}))\in\mathcal{T}_{\epsilon'}^{(n)},
\end{align}
where $\hat{m}_{j-2,2}$ was decoded in block~$j$, and $\hat{m}_{j-1,2}$ was decoded in block~$j+1$. At the end of block~$j+2$, the decoder receiver~2 then finds the unique message $\hat{m}_{j,2}$ s.t.,
\begin{multline}
	(x^n(\hat{m}_{j,2}\vert\hat{m}_{j-2,2}\vert\hat{m}_{j-3,2}),x_1^n(\hat{m}_{j-2,2}\vert\hat{m}_{j-3,2}),x_3^n(\hat{l}_{j-1,3}),\tilde{y}_3^n(\hat{k}_{j,3}\vert\hat{l}_{j-1,3}),\\
	x_2^n(l_{j-1,2}\vert\hat{m}_{j-3,2}),y_2^n(j),u^n(\hat{m}_{j-3,2}))\in\mathcal{T}_{\epsilon}^{(n)},
\end{multline}
for some $\hat{k}_{j,3}\in\mathcal{B}(\hat{l}_{j,3})$, and,
\begin{align}
	(x_1^n(\hat{m}_{j,2}\vert\hat{m}_{j-1,2}),x_2^n(l_{j+1,2}\vert\hat{m}_{j-1,2}),y_2^n(j+2),u^n(\hat{m}_{j-1,2}))\in\mathcal{T}_{\epsilon}^{(n)}
\end{align}
simultaneously.

\emph{Backward decoding at receiver~3.} After all $b$~blocks are received the decoder receiver~3 realizes a backward decoding. For $j=b-3,b-4,\ldots,1$, the decoder receiver~3 finds the unique message $\hat{m}_{j,3}$ s.t.\ there exist a $l_{j+2,2}\in[1:2^{nR_2}]$ s.t.,
\begin{align}
	(x^n(\hat{m}_{j+3,3}\vert\hat{m}_{j+1,3}\vert\hat{m}_{j,3}),x_1^n(\hat{m}_{j+1,3}\vert\hat{m}_{j,3}),x_2^n(l_{j+2,2}\vert\hat{m}_{j,3}),x_3^n(l_{j+2,3}),y_3^n(j+3),u^n(\hat{m}_{j,3}))\in\mathcal{T}_{\epsilon}^{(n)},
\end{align}
successively with the initial conditions $\hat{m}_{b-2,3}=\hat{m}_{b-1,3}=\hat{m}_{b,3}=1$. If there is more than one such index, it selects one of them uniformly at random.

The probability of decoding error is analyzed at the decoder receivers~1, 2, and~3, for the message $M_j$ averaged over codebooks.

\emph{Analysis of the probability of error at receiver~1.} Assume without loss of generality that $M_{j-3}=M_{j-2}=M_j=1$ and let $L_{j-1,2},L_{j,2},K_{j,2},L_{j-1,3},L_{j,3},K_{j,3}$ denote the indices chosen by the relays receiver~2 and receiver~3 in blocks~$j$ and~$j+1$. Then, the decoder receiver~1 makes an error only if one or more of the following events occur,
\begin{align}
	\mathcal{E}_{(1)1,2,3}^{(2,3)}&(j-3)=\left\{\hat{M}_{j-3,1}\neq1\right\}\textrm{ and }\mathcal{E}_{(1)1,2,3}^{(2,3)}(j-2)\label{eq:Analysis of the probability of error at receiver 1 - 1}\\
	\tilde{\mathcal{E}}_{(2)}(j)=&\Big\{(Y_2^n(j),\tilde{Y}_2^n(k_{j,2}\vert L_{j-1,2}\vert\hat{M}_{j-3,2}),X_2^n(L_{j-1,2}\vert\hat{M}_{j-3,2}),U^n(\hat{M}_{j-3,2}))\not\in\mathcal{T}_{\epsilon'}^{(n)},\nonumber\\
	&\forall k_{j,2}\in[1:2^{n\tilde{R}_2}]\Big\}\label{eq:Analysis of the probability of error at receiver 1 - 2}
\end{align}

\begin{align}
	\tilde{\mathcal{E}}_{(3)}(j)=&\Big\{(Y_3^n(j),\tilde{Y}_3^n(k_{j,3}\vert L_{j-1,3}),X_3^n(L_{j-1,3}))\not\in\mathcal{T}_{\epsilon'}^{(n)},\ \forall k_{j,3}\in[1:2^{n\tilde{R}_3}]\Big\}\label{eq:Analysis of the probability of error at receiver 1 - 3}\\
	\mathcal{E}_{(1)1}(j)=&\left\{\hat{L}_{j,2}\neq L_{j,2}\right\}\textrm{ and }\mathcal{E}_{(1)1}(j-1)\label{eq:Analysis of the probability of error at receiver 1 - 4}\\
	\mathcal{E}_{(1)2}(j)=&\left\{\hat{L}_{j,3}\neq L_{j,3}\right\}\textrm{ and }\mathcal{E}_{(1)2}(j-1)\label{eq:Analysis of the probability of error at receiver 1 - 5}\\
	\mathcal{E}_{(1)3}(j)=&\Big\{(X^n(1\vert\hat{M}_{j-2,1}\vert\hat{M}_{j-3,1}),X_2^n(\hat{L}_{j-1,2}\vert\hat{M}_{j-3,1}),\tilde{Y}_2^n(K_{j,2}\vert\hat{L}_{j-1,2}\vert\hat{M}_{j-3,1}),\nonumber\\
	&X_3^n(\hat{L}_{j-1,3}),\tilde{Y}_3^n(K_{j,3}\vert\hat{L}_{j-1,3}),X_1^n(\hat{M}_{j-2,1}\vert\hat{M}_{j-3,1}),Y_1^n(j),U^n(\hat{M}_{j-3,1}))\not\in\mathcal{T}_{\epsilon}^{(n)}\Big\}\label{eq:Analysis of the probability of error at receiver 1 - 6}\\
	\mathcal{E}_{(1)4}(j)=&\Big\{(X^n(m_{j,1}\vert\hat{M}_{j-2,1}\vert\hat{M}_{j-3,1}),X_2^n(\hat{L}_{j-1,2}\vert\hat{M}_{j-3,1}),\tilde{Y}_2^n(K_{j,2}\vert\hat{L}_{j-1,2}\vert\hat{M}_{j-3,1}),\nonumber\\
	&X_3^n(\hat{L}_{j-1,3}),\tilde{Y}_3^n(K_{j,3}\vert\hat{L}_{j-1,3}),X_1^n(\hat{M}_{j-2,1}\vert\hat{M}_{j-3,1}),Y_1^n(j),U^n(\hat{M}_{j-3,1}))\in\mathcal{T}_{\epsilon}^{(n)}\nonumber\\
	&\textrm{for some }m_{j,1}\neq1\Big\}\label{eq:Analysis of the probability of error at receiver 1 - 7}\\
	\mathcal{E}_{(1)5}(j)=&\Big\{(X^n(m_{j,1}\vert\hat{M}_{j-2,1}\vert\hat{M}_{j-3,1}),X_2^n(\hat{L}_{j-1,2}\vert\hat{M}_{j-3,1}),\tilde{Y}_2^n(\hat{k}_{j,2}\vert\hat{L}_{j-1,2}\vert\hat{M}_{j-3,1}),\nonumber\\
	&X_3^n(\hat{L}_{j-1,3}),\tilde{Y}_3^n(K_{j,3}\vert\hat{L}_{j-1,3}),X_1^n(\hat{M}_{j-2,1}\vert\hat{M}_{j-3,1}),Y_1^n(j),U^n(\hat{M}_{j-3,1}))\in\mathcal{T}_{\epsilon}^{(n)}\nonumber\\
	&\textrm{for some }\hat{k}_{j,2}\in\mathcal{B}(\hat{L}_{j,2}),\hat{k}_{j,2}\neq K_{j,2},m_{j,1}\neq1\Big\}\label{eq:Analysis of the probability of error at receiver 1 - 8}\\
	\mathcal{E}_{(1)6}(j)=&\Big\{(X^n(m_{j,1}\vert\hat{M}_{j-2,1}\vert\hat{M}_{j-3,1}),X_2^n(\hat{L}_{j-1,2}\vert\hat{M}_{j-3,1}),\tilde{Y}_2^n(K_{j,2}\vert\hat{L}_{j-1,2}\vert\hat{M}_{j-3,1}),\nonumber\\
	&X_3^n(\hat{L}_{j-1,3}),\tilde{Y}_3^n(\hat{k}_{j,3}\vert\hat{L}_{j-1,3}),X_1^n(\hat{M}_{j-2,1}\vert\hat{M}_{j-3,1}),Y_1^n(j),U^n(\hat{M}_{j-3,1}))\in\mathcal{T}_{\epsilon}^{(n)}\nonumber\\
	&\textrm{for some }\hat{k}_{j,3}\in\mathcal{B}(\hat{L}_{j,3}),\hat{k}_{j,3}\neq K_{j,3},m_{j,1}\neq1\Big\}\label{eq:Analysis of the probability of error at receiver 1 - 9}\\
	\mathcal{E}_{(1)7}(j)=&\Big\{(X^n(m_{j,1}\vert\hat{M}_{j-2,1}\vert\hat{M}_{j-3,1}),X_2^n(\hat{L}_{j-1,2}\vert\hat{M}_{j-3,1}),\tilde{Y}_2^n(\hat{k}_{j,2}\vert\hat{L}_{j-1,2}\vert\hat{M}_{j-3,1}),\nonumber\\
	&X_3^n(\hat{L}_{j-1,3}),\tilde{Y}_3^n(\hat{k}_{j,3}\vert\hat{L}_{j-1,3}),X_1^n(\hat{M}_{j-2,1}\vert\hat{M}_{j-3,1}),Y_1^n(j),U^n(\hat{M}_{j-3,1}))\in\mathcal{T}_{\epsilon}^{(n)}\nonumber\\
	&\textrm{for some }\hat{k}_{j,2}\in\mathcal{B}(\hat{L}_{j,2}),\hat{k}_{j,2}\neq K_{j,2},\hat{k}_{j,3}\in\mathcal{B}(\hat{L}_{j,3}),\hat{k}_{j,3}\neq K_{j,3},m_{j,1}\neq1\Big\}.\label{eq:Analysis of the probability of error at receiver 1 - 10}
\end{align}

\emph{Analysis of the probability of error at receiver~2.} Assume without loss of generality that $M_{j-3}=M_{j-2}=M_{j-1}=M_j=1$ and let $\hat{M}_{j-3,1},\hat{M}_{j-2,1},\hat{M}_{j-1,1},\hat{M}_{j,1}$ denote the indices chosen by the relay receiver~1 in blocks~$j-1,j,j+1$ and~$j+2$, and $\hat{M}_{j-3,2},\hat{M}_{j-2,2},\hat{M}_{j-1,2}$ be the relay estimate of $\hat{M}_{j-3,1},\hat{M}_{j-2,1},\hat{M}_{j-1,1}$ at the decoder receiver~2, and let $L_{j-1,3},L_{j,3},K_{j,3}$ denote the indices chosen by the relay receiver~3 in blocks~$j$ and~$j+1$, and $\hat{L}_{j-1,3},\hat{L}_{j,3}$ be the relay estimates of $L_{j-1,3},L_{j,3}$ at the decoder receiver~2. Then, the decoder receiver~2 makes an error only if one or more of the following events occur,
\begin{align}
	\mathcal{E}_{(1)1,2,3}^{(2,3)}&(j-3)=\left\{\hat{M}_{j-3,1}\neq1\right\},\ \mathcal{E}_{(1)1,2,3}^{(2,3)}(j-2),\ \mathcal{E}_{(1)1,2,3}^{(2,3)}(j-1),\textrm{ and }\mathcal{E}_{(1)1,2,3}^{(2,3)}(j)\label{eq:Analysis of the probability of error at receiver 2 - 1}
\end{align}

\begin{align}
	\mathcal{E}_{(2)1,2,3}^{(2,3)}&(j-3)=\left\{\hat{M}_{j-3,2}\neq1\right\},\ \mathcal{E}_{(2)1,2,3}^{(2,3)}(j-2),\textrm{ and }\mathcal{E}_{(2)1,2,3}^{(2,3)}(j-1)\label{eq:Analysis of the probability of error at receiver 2 - 2}\\
	\tilde{\mathcal{E}}_{(3)}(j)=&\Big\{(Y_3^n(j),\tilde{Y}_3^n(k_{j,3}\vert L_{j-1,3}),X_3^n(L_{j-1,3}))\not\in\mathcal{T}_{\epsilon'}^{(n)},\ \forall k_{j,3}\in[1:2^{n\tilde{R}_3}]\Big\}\label{eq:Analysis of the probability of error at receiver 2 - 3}\\
	\mathcal{E}_{(2)1}(j)=&\left\{\hat{L}_{j,3}\neq L_{j,3}\right\}\textrm{ and }\mathcal{E}_{(2)1}(j-1)\label{eq:Analysis of the probability of error at receiver 2 - 4}\\
	\mathcal{E}_{(2)2}(j)=&\Big\{(X^n(\hat{M}_{j,2}\vert\hat{M}_{j-2,2}\vert\hat{M}_{j-3,2}),X_1^n(\hat{M}_{j-2,2}\vert\hat{M}_{j-3,2}),X_2^n(L_{j-1,2}\vert\hat{M}_{j-3,2}),\nonumber\\
	&X_3^n(\hat{L}_{j-1,3}),\tilde{Y}_3^n(K_{j,3}\vert\hat{L}_{j-1,3}),Y_2^n(j),U^n(\hat{M}_{j-3,2}))\not\in\mathcal{T}_{\epsilon}^{(n)}\nonumber\\
	&\textrm{or }(X_1^n(\hat{M}_{j,2}\vert\hat{M}_{j-1,2}),X_2^n(L_{j+1,2}\vert\hat{M}_{j-1,2}),Y_2^n(j+2),U^n(\hat{M}_{j-1,2}))\not\in\mathcal{T}_{\epsilon}^{(n)}\Big\}\label{eq:Analysis of the probability of error at receiver 2 - 5}\\
	\mathcal{E}_{(2)3}(j)=&\Big\{(X^n(m_{j,2}\vert\hat{M}_{j-2,2}\vert\hat{M}_{j-3,2}),X_1^n(\hat{M}_{j-2,2}\vert\hat{M}_{j-3,2}),X_2^n(L_{j-1,2}\vert\hat{M}_{j-3,2}),\nonumber\\
	&X_3^n(\hat{L}_{j-1,3}),\tilde{Y}_3^n(K_{j,3}\vert\hat{L}_{j-1,3}),Y_2^n(j),U^n(\hat{M}_{j-3,2}))\in\mathcal{T}_{\epsilon}^{(n)}\nonumber\\
	&\textrm{and }(X_1^n(m_{j,2}\vert\hat{M}_{j-1,2}),X_2^n(L_{j+1,2}\vert\hat{M}_{j-1,2}),Y_2^n(j+2),U^n(\hat{M}_{j-1,2}))\in\mathcal{T}_{\epsilon}^{(n)}\nonumber\\
	&\textrm{for some }m_{j,2}\neq\hat{M}_{j,1}\Big\}\label{eq:Analysis of the probability of error at receiver 2 - 6}\\
	\mathcal{E}_{(2)4}(j)=&\Big\{(X^n(m_{j,2}\vert\hat{M}_{j-2,2}\vert\hat{M}_{j-3,2}),X_1^n(\hat{M}_{j-2,2}\vert\hat{M}_{j-3,2}),X_2^n(L_{j-1,2}\vert\hat{M}_{j-3,2}),\nonumber\\
	&X_3^n(\hat{L}_{j-1,3}),\tilde{Y}_3^n(\hat{k}_{j,3}\vert\hat{L}_{j-1,3}),Y_2^n(j),U^n(\hat{M}_{j-3,2}))\in\mathcal{T}_{\epsilon}^{(n)}\nonumber\\
	&\textrm{and }(X_1^n(m_{j,2}\vert\hat{M}_{j-1,2}),X_2^n(L_{j+1,2}\vert\hat{M}_{j-1,2}),Y_2^n(j+2),U^n(\hat{M}_{j-1,2}))\in\mathcal{T}_{\epsilon}^{(n)}\nonumber\\
	&\textrm{for some }\hat{k}_{j,3}\in\mathcal{B}(\hat{L}_{j,3}),\hat{k}_{j,3}\neq K_{j,3},m_{j,2}\neq\hat{M}_{j,1}\Big\}.\label{eq:Analysis of the probability of error at receiver 2 - 7}
\end{align}

\emph{Analysis of the probability of error at receiver~3.} Assume without loss of generality that $M_j=M_{j+1}=M_{j+3}=1$ and let $\hat{M}_{j,1},\hat{M}_{j+1,1}$ denote the indices chosen by the relay receiver~1 in block~$j+3$, $L_{j+2,2}=1,\hat{M}_{j,2}$ denote the indices chosen by the relay receiver~2 in block~$j+3$, and $\hat{M}_{j+1,3},\hat{M}_{j+3,3}$ be the relay estimate of $M_{j+1},M_{j+3}$ at the decoder receiver~3, and $L_{j+2,3}=1$ denote the index chosen by the relay receiver~3 in block~$j+3$. Then, the decoder receiver~3 makes an error only if one or more of the following events occur,
\begin{align}
	\mathcal{E}_{(1)1,2,3}^{(2,3)}&(j)=\left\{\hat{M}_{j,1}\neq1\right\}\textrm{ and }\mathcal{E}_{(1)1,2,3}^{(2,3)}(j+1)=\left\{\hat{M}_{j+1,1}\neq1\right\}\label{eq:Analysis of the probability of error at receiver 3 - 1}\\
	\mathcal{E}_{(2)1,2,3}^{(2,3)}&(j)=\left\{\hat{M}_{j,2}\neq1\right\}\label{eq:Analysis of the probability of error at receiver 3 - 2}\\
	\mathcal{E}_{(3)1,2,3}^{(2,3)}&(j+1)=\left\{\hat{M}_{j+1,3}\neq1\right\}\textrm{ and }\mathcal{E}_{(3)1,2,3}^{(2,3)}(j+3)=\left\{\hat{M}_{j+3,3}\neq1\right\}\label{eq:Analysis of the probability of error at receiver 3 - 3}
\end{align}

\begin{align}
	\mathcal{E}_{(3)1}(j)=&\Big\{(X^n(\hat{M}_{j+3,3}\vert\hat{M}_{j+1,3}\vert\hat{M}_{j,3}),X_1^n(\hat{M}_{j+1,3}\vert\hat{M}_{j,3}),X_2^n(l_{j+2,2}\vert\hat{M}_{j,3}),X_3^n(L_{j+2,3}),\nonumber\\
	&Y_3^n(j+3),U^n(\hat{M}_{j,3}))\not\in\mathcal{T}_{\epsilon}^{(n)},\ \forall l_{j+2,2}\in[1:2^{nR_2}]\Big\}\label{eq:Analysis of the probability of error at receiver 3 - 4}\\
	\mathcal{E}_{(3)2}(j)=&\Big\{X^n(\hat{M}_{j+3,3}\vert\hat{M}_{j+1,3}\vert m_{j,3}),X_1^n(\hat{M}_{j+1,3}\vert m_{j,3}),X_2^n(l_{j+2,2}\vert m_{j,3}),X_3^n(L_{j+2,3}),\nonumber\\
	&Y_3^n(j+3),U^n(m_{j,3}))\in\mathcal{T}_{\epsilon}^{(n)}\textrm{ for some }m_{j,3}\neq1,l_{j+2,2}\in[1:2^{nR_2}]\Big\}.\label{eq:Analysis of the probability of error at receiver 3 - 5}
\end{align}
Note that receiver~3 does not decode $l_{j+2,2}$ since it contains information about $m_{j+2}$, which has already been decoded in the backward decoding procedure, and so it does not bring any new information. Thus, either $\hat{m}_{j,3}=m_j$ and receiver~3 only needs to find a satellite index $l_{j+2,2}$ so that all the sequences considered are typical, or $\hat{m}_{j,3}\neq m_j$ and the cloud center selected at receiver~3 for $x_2^n(l_{j+2,2}\vert\hat{m}_{j,3})$ is not the right one, so the index has no further impact on the probability of error since only the $m_j$ has to be correctly decoded at each receiver. Moreover, as previously underlined, if sliding window decoding is implemented, the scheme ends for a given message $m_j$ in the block~$j+4$, giving a latency of 3~blocks. We used backward decoding at receiver~3 to ease the error events provided.

Due to space restrictions, the details of the formal proof are omitted. By induction, the probability of error of the terms~\eqref{eq:Analysis of the probability of error at receiver 1 - 1}, \eqref{eq:Analysis of the probability of error at receiver 2 - 1}, \eqref{eq:Analysis of the probability of error at receiver 2 - 2}, and~\eqref{eq:Analysis of the probability of error at receiver 3 - 1}-\eqref{eq:Analysis of the probability of error at receiver 3 - 3} tend to zero as $n\rightarrow\infty$ for every $j\in[1:b-3]$, if the bounds on the probability of error of the remaining terms are satisfied. Applying 1)~the union of events bound, the independence of the codebooks, the law of large numbers, the conditional typicality lemma~\cite[Sec.~2.5]{book3}, the joint typicality lemma~\cite[Sec.~2.5.1]{book3}, the packing lemma~\cite[Lem.~3.1, Sec~3.2]{book3}, the covering lemma~\cite[Lem.~3.3, Sec~3.7]{book3}, the lemma 11.1~\cite[Sec.~11.3.1]{book3}, the chain rule, and the Fourier-Motzkin elimination procedure~\cite[Appx.~D]{book3} on~\eqref{eq:Analysis of the probability of error at receiver 1 - 2}-\eqref{eq:Analysis of the probability of error at receiver 1 - 10}, \eqref{eq:Analysis of the probability of error at receiver 2 - 3}-\eqref{eq:Analysis of the probability of error at receiver 2 - 7}, 2)~the union of events bound, the independence of the codebooks, the law of large numbers, and the packing lemma on~\eqref{eq:Analysis of the probability of error at receiver 3 - 4}, and~\eqref{eq:Analysis of the probability of error at receiver 3 - 5}, 3)~combining the resulting bounds, 4)~taking the limit over $n$, and 5)~maximizing over the six sub-strategies gives the result in Prop.~\ref{prop:3FC}.

\section{Special case: 3PC}
\label{appx:Special case: 3PC}

The bounds of $\textrm{STG}_{1,2,3}^{(2,3)}$ can be specialized to $\textrm{STG}_{1,2,3}^{(2)}$ by restricting $\tilde{Y}_3$ to be independent of $(X_3,Y_3)$ and setting $X_3$ to be a function of a constant. Thus, the probability distribution
\begin{align}
	p(u)p(x_1\vert u)p(x\vert x_1)p(x_2\vert u)p(x_3)p(\tilde{y}_2\vert x_2,y_2)p(\tilde{y}_3\vert x_3,y_3)
\end{align}
becomes
\begin{align}
	p(u)p(x_1\vert u)p(x\vert x_1)p(x_2\vert u)p(x_3)p(\tilde{y}_2\vert x_2,y_2).
\end{align}
After 1)~applying the simplification, 2)~removing the bounds that appear twice and using the chain rule, and 3)~noticing that in the remaining bounds, two bounds are the average of respectively two other ones and thus are never active, since $\forall a,b\in\mathds{R}^+,\ \frac{1}{2}(a+b)\geq\min\{a,b\}$, one can get the bounds in Coro.~\ref{coro:3PC}.

\section{Special case: 3FC in the Gaussian case}
\label{appx:Special case: 3FC in the Gaussian case}

The bounds of $\textrm{STG}_{1,2,3}^{(2,3)}$ can be specialized to $\textrm{STG}_{1,2,3}^{(2,3)\textrm{Gauss}}$ as follows. Assume that $U\sim\mathcal{CN}(0,\sigma_U^2)$, with $\sigma_U^2=\mathbb{E}[UU^*]=P\rho_U,\ 0\leq\rho_U\leq1$. Assume that $X_1=X_1'+A_1U\sim\mathcal{CN}(0,\sigma_{X_1}^2)$, with $\sigma_{X_1}^2=\mathbb{E}[X_1X_1^*]=P(\rho_{X_1'}+\rho_U\rho_{A_1}^2),\ 0\leq\rho_{X_1'}\leq1,\ 0\leq\rho_{A_1}\leq1,\ \theta_{A_1}\in[0,2\pi)$, with correlation coefficient $Q_{U,X_1}=\mathbb{E}[UX_1^*]=P\rho_U\rho_{A_1}e^{-\textrm{j}\theta_{A_1}}$. Assume that $X=X'+A_2X_1\sim\mathcal{CN}(0,\sigma_{X}^2)$, with $\sigma_{X}^2=\mathbb{E}[XX^*]=P(\rho_{X'}+\rho_{X_1'}\rho_{A_2}^2+\rho_U\rho_{A_1}^2\rho_{A_2}^2),\ 0\leq\rho_{X'}\leq1,\ 0\leq\rho_{A_2}\leq1,\ \theta_{A_2}\in[0,2\pi)$, with correlation coefficients $Q_{U,X}=\mathbb{E}[UX^*]=P\rho_U\rho_{A_1}\rho_{A_2}e^{-\textrm{j}(\theta_{A_1}+\theta_{A_2})}$, and $Q_{X,X_1}=\mathbb{E}[XX_1^*]=P(\rho_{X_1'}\rho_{A_2}+\rho_U\rho_{A_1}^2\rho_{A_2})e^{\textrm{j}\theta_{A_2}}$. Assume that $X_2=X_2'+B_1U\sim\mathcal{CN}(0,\sigma_{X_2}^2)$, with $\sigma_{X_2}^2=\mathbb{E}[X_2X_2^*]=P(\rho_{X_2'}+\rho_U\rho_{B_1}^2),\ 0\leq\rho_{X_2'}\leq1,\ 0\leq\rho_{B_1}\leq1,\ \theta_{B_1}\in[0,2\pi)$, with correlation coefficients $Q_{U,X_2}=\mathbb{E}[UX_2^*]=P\rho_U\rho_{B_1}e^{-\textrm{j}\theta_{B_1}}$, $Q_{X,X_2}=\mathbb{E}[XX_2^*]=P\rho_U\rho_{A_1}\rho_{A_2}\rho_{B_1}e^{\textrm{j}(\theta_{A_1}+\theta_{A_2}-\theta_{B_1})}$, and $Q_{X_1,X_2}=\mathbb{E}[X_1X_2^*]=P\rho_U\rho_{A_1}\rho_{B_1}e^{\textrm{j}(\theta_{A_1}-\theta_{B_1})}$. Assume that $X_3\sim\mathcal{CN}(0,\sigma_{X_3}^2)$, with $\sigma_{X_3}^2=\mathbb{E}[X_3X_3^*]=P\rho_{X_3},\ 0\leq\rho_{X_3}\leq1$. The AWGN $Z_k\sim\mathcal{CN}(0,1),\ k\in[1:3]$, are independent and i.i.d.\ over time and for every block. The quantization random variables are defined as $\tilde{Y}_k=Y_k+\tilde{Z}_k,\ \tilde{Z}_k\sim\mathcal{CN}(0,\Delta_k),\ k\in[2:3]$, and the $\tilde{Z}_k$ are independent of everything else. Giving the covariance matrix,
\begin{align}
	&\mat{\Sigma_{1,2,3}^{(2,3)}}\\
	&=
	\left[
		\begin{array}{ccccc}
			\sigma_U^2&Q_{U,X}&Q_{U,X_1}&Q_{U,X_2}&0\\
			Q_{U,X}^*&\sigma_X^2&Q_{X,X_1}&Q_{X,X_2}&0\\
			Q_{U,X_1}^*&Q_{X,X_1}^*&\sigma_{X_1}^2&Q_{X_1,X_2}&0\\
			Q_{U,X_2}^*&Q_{X,X_2}^*&Q_{X_1,X_2}^*&\sigma_{X_2}^2&0\\
			0&0&0&0&\sigma_{X_3}^2
		\end{array}
	\right]
	\\
	&=
	\begin{bsmallmatrix}
		P\rho_U&P\rho_U\rho_{A_1}\rho_{A_2}e^{-\textrm{j}(\theta_{A_1}+\theta_{A_2})}&P\rho_U\rho_{A_1}e^{-\textrm{j}\theta_{A_1}}&P\rho_U\rho_{B_1}e^{-\textrm{j}\theta_{B_1}}&0\\
		Q_{U,X}^*&P(\rho_{X'}+\rho_{X_1'}\rho_{A_2}^2+\rho_U\rho_{A_1}^2\rho_{A_2}^2)&P(\rho_{X_1'}\rho_{A_2}+\rho_U\rho_{A_1}^2\rho_{A_2})e^{\textrm{j}\theta_{A_2}}&P\rho_U\rho_{A_1}\rho_{A_2}\rho_{B_1}e^{\textrm{j}(\theta_{A_1}+\theta_{A_2}-\theta_{B_1})}&0\\
		Q_{U,X_1}^*&Q_{X,X_1}^*&P(\rho_{X_1'}+\rho_U\rho_{A_1}^2)&P\rho_U\rho_{A_1}\rho_{B_1}e^{\textrm{j}(\theta_{A_1}-\theta_{B_1})}&0\\
		Q_{U,X_2}^*&Q_{X,X_2}^*&Q_{X_1,X_2}^*&P(\rho_{X_2'}+\rho_U\rho_{B_1}^2)&0\\
		0&0&0&0&P\rho_{X_3}
	\end{bsmallmatrix},\label{eq:covariance matrix}
\end{align}
which is positive semi-definite, and $\mat{\Sigma_{1,2,3}^{(2,3)}}\preceq P\mat{I_5}$, thus all the diagonal elements are smaller or equal to $P$. The latter constraint is used for comparison with the other schemes, however, it can be noticed that this is more strict that $\Tr(\mat{\Sigma_{1,2,3}^{(2)}})\leq5P$. It can be noticed that without loss of generality, $U$ and $X_3$ do not require a phase under this setting. Note that diagonal elements of $\mat{\Sigma_{1,2,3}^{(2,3)}}$ modulate the power allocation dedicated to each layer of the superposition of CFs and DFs. By applying the $\log\det(\cdot)$ on the bounds, this leads to the expression presented in Coro.~\ref{coro:3FC Gaussian}. In a similar manner, one can derive the $\textrm{STG}_{1,2,3}^{(2)\textrm{Gauss}}$ for the 3PC scheme.

\section{Special case: 2RC}
\label{appx:Special case: 2RC}

The bounds of $\textrm{STG}_{1,2,3}^{(2)}$ can be specialized to $\textrm{STG}_{1,2}^{(2)}$ by restricting $Y_3$ to be independent of $(U,X,X_1,X_2,X_3)$, by not requiring receiver~3 to decode the common message anymore, and by setting $U$ to be a function of a constant. Thus, the probability distribution
\begin{align}
	p(u)p(x_1\vert u)p(x\vert x_1)p(x_2\vert u)p(x_3)p(\tilde{y}_2\vert x_2,y_2)
\end{align}
becomes
\begin{align}
	p(x,x_1)p(x_2)p(x_3)p(\tilde{y}_2\vert x_2,y_2).
\end{align}
After applying the simplification, one can get the bounds presented in~\cite{ownpublications1}, where there are further specialized to the orthogonal case, the SISO Gaussian case and the MISO Gaussian case.

\bibliographystyle{IEEEtran}
\bibliography{IEEEabrv,Biblio}

\begin{thebibliography}{10}
\providecommand{\url}[1]{#1}
\csname url@samestyle\endcsname
\providecommand{\newblock}{\relax}
\providecommand{\bibinfo}[2]{#2}
\providecommand{\BIBentrySTDinterwordspacing}{\spaceskip=0pt\relax}
\providecommand{\BIBentryALTinterwordstretchfactor}{4}
\providecommand{\BIBentryALTinterwordspacing}{\spaceskip=\fontdimen2\font plus
\BIBentryALTinterwordstretchfactor\fontdimen3\font minus
  \fontdimen4\font\relax}
\providecommand{\BIBforeignlanguage}[2]{{%
\expandafter\ifx\csname l@#1\endcsname\relax
\typeout{** WARNING: IEEEtran.bst: No hyphenation pattern has been}%
\typeout{** loaded for the language `#1'. Using the pattern for}%
\typeout{** the default language instead.}%
\else
\language=\csname l@#1\endcsname
\fi
#2}}
\providecommand{\BIBdecl}{\relax}
\BIBdecl

\bibitem{article41}
X.~Lin, J.~G. Andrews, A.~Ghosh, and R.~Ratasuk, ``An overview on {3GPP}
  device-to-device proximity services,'' \emph{IEEE Commun. Mag.}, vol.~52,
  no.~4, pp. 40--48, Apr. 2014.

\bibitem{unpublished5}
Z.~Xiang, M.~Tao, and X.~Wang, ``Massive {MIMO} multicasting in noncooperative
  cellular networks,'' 2014, arXiv:1312.1134.

\bibitem{misc2}
{3GPP: Technical Specification Group on Service and System Aspects (3GPP TSG
  SA)}, ``{Feasibility study for proximity services (ProSe) (release 12)},''
  Jun. 2013, {TR 22.803, ver. 12.2.0}.

\bibitem{misc3}
{Wi-Fi Alliance}, ``{Wi-Fi peer-to-peer (P2P) technical specification},'' 2016,
  {ver. 1.5}.

\bibitem{inproceedings3}
S.~Draper, B.~Frey, and F.~Kschischang, ``Interactive decoding of a broadcast
  message,'' in \emph{Proc. 41st Allerton Conf. on Commun., Control, and
  Comput.}, 2003.

\bibitem{article26}
E.~van~der Meulen, ``Three-terminal communication channels,'' \emph{Adv. Appl.
  Prob.}, vol.~3, no.~1, pp. 120--154, Spring 1971.

\bibitem{article30}
T.~Cover and A.~El~Gamal, ``Capacity theorems for the relay channel,''
  \emph{IEEE Trans. Inf. Theory}, vol.~25, no.~5, pp. 572--584, Sep. 1979.

\bibitem{article23}
G.~Kramer, M.~Gastpar, and P.~Gupta, ``Cooperative strategies and capacity
  theorems for relay networks,'' \emph{IEEE Trans. Inf. Theory}, vol.~51,
  no.~9, pp. 3037--3063, Sep. 2005.

\bibitem{inproceedings13}
S.~H. Lim, Y.-H. Kim, A.~El~Gamal, and S.-Y. Chung, ``Noisy network coding,''
  in \emph{Proc. Inf. Theory Workshop (ITW)}, 2010, pp. 1--5.

\bibitem{article32}
------, ``Noisy network coding,'' \emph{IEEE Trans. Inf. Theory}, vol.~57,
  no.~5, pp. 3132--3152, May 2011.

\bibitem{article33}
J.~Hou and G.~Kramer, ``Short message noisy network coding with a
  decode-forward option,'' \emph{IEEE Trans. Inf. Theory}, vol.~62, no.~1, pp.
  89--107, Jan. 2016.

\bibitem{article31}
A.~Behboodi and P.~Piantanida, ``Cooperative strategies for simultaneous and
  broadcast relay channels,'' \emph{IEEE Trans. Inf. Theory}, vol.~59, no.~3,
  pp. 1417--1443, Mar. 2013.

\bibitem{unpublished10}
I.~Mari\'c and D.~Hui, ``Short message noisy network coding with rate
  splitting,'' 2014, arXiv:1404.0061.

\bibitem{article14}
R.~Dabora and S.~D. Servetto, ``Broadcast channels with cooperating decoders,''
  \emph{IEEE Trans. Inf. Theory}, vol.~52, no.~12, pp. 5438--5454, Dec. 2006.

\bibitem{article19}
Y.~Liang and V.~V. Veeravalli, ``Cooperative relay broadcast channels,''
  \emph{IEEE Trans. Inf. Theory}, vol.~53, no.~3, pp. 900--928, Mar. 2007.

\bibitem{article21}
Y.~Liang and G.~Kramer, ``Rate regions for relay broadcast channels,''
  \emph{IEEE Trans. Inf. Theory}, vol.~53, no.~10, pp. 3517--3535, Oct. 2007.

\bibitem{ownpublications1}
V.~Exposito, S.~Yang, and N.~Gresset, ``A two-round interactive receiver
  cooperation scheme for multicast channels,'' in \emph{Proc. 54th Allerton
  Conf. on Commun., Control, and Comput.}, 2016, to appear.

\bibitem{inproceedings14}
J.~Du, W.~Zhu, J.~Xu, Z.~Li, and H.~Wang, ``A compressed {HARQ} feedback for
  device-to-device multicast communications,'' in \emph{Proc. Trans. Veh.
  Technol. (VTC Fall)}, 2012, pp. 1--5.

\bibitem{article35}
A.~Asadi, Q.~Wang, and V.~Mancuso, ``A survey on device-to-device communication
  in cellular networks,'' \emph{IEEE Commun. Surveys Tuts.}, vol.~16, no.~4,
  pp. 1801--1819, Fourthquarter 2014.

\bibitem{unpublished6}
C.~Karakus and S.~Diggavi, ``Enhancing multiuser {MIMO} through opportunistic
  {D2D} cooperation,'' 2016, arXiv:1604.06151.

\bibitem{article20}
A.~Steiner, A.~Sanderovich, and S.~Shamai~(Shitz), ``Broadcast cooperation
  strategies for two colocated users,'' \emph{IEEE Trans. Inf. Theory},
  vol.~53, no.~10, pp. 3394--3412, Oct. 2007.

\bibitem{unpublished7}
X.~Lin, R.~Ratasuk, A.~Ghosh, and J.~G. Andrews, ``Modeling, analysis and
  optimization of multicast device-to-device transmission,'' 2013,
  arXiv:1309.1518.

\bibitem{unpublished8}
X.~Lin, J.~G. Andrews, and A.~Ghosh, ``Spectrum sharing for device-to-device
  communication in cellular networks,'' 2013, arXiv:1305.4219.

\bibitem{article37}
A.~Khisti, U.~Erez, and G.~Wornell, ``Fundamental limits and scaling behavior
  of cooperative multicasting in wireless networks,'' \emph{IEEE Trans. Inf.
  Theory}, vol.~52, no.~6, pp. 2762--2770, Jun. 2006.

\bibitem{article38}
I.~Mari\'c and R.~D. Yates, ``Cooperative multicast for maximum network
  lifetime,'' \emph{IEEE J. Sel. Areas Commun.}, vol.~23, no.~1, pp. 127--135,
  Jan. 2005.

\bibitem{inproceedings16}
M.~Duarte and A.~Sabharwal, ``Full-duplex wireless communications using
  off-the-shelf radios: Feasibility and first results,'' in \emph{Proc. 44th
  Asilomar Conf. on Signals, Syst. and Comput. (ACSSC)}, 2010, pp. 1558--1562.

\bibitem{article43}
M.~Duarte, C.~Dick, and A.~Sabharwal, ``Experiment-driven characterization of
  full-duplex wireless systems,'' \emph{IEEE Trans. Wireless Commun.}, vol.~11,
  no.~12, pp. 4296--4307, Dec. 2012.

\bibitem{inproceedings15}
N.~Shende, O.~Gurbuz, and E.~Erkip, ``Half-duplex or full-duplex relaying: A
  capacity analysis under self-interference,'' in \emph{Proc. 47th Annual Conf.
  on Information Sciences and Syst. (CISS)}, 2013, pp. 1--6.

\bibitem{ownpublications2}
V.~Exposito, S.~Yang, and N.~Gresset, ``Multicast channel communication with
  interactive receiver cooperation over orthogonal links,'' in \emph{Proc.
  Internat. Conf. on Commun. (ICC)}, 2017, accepted.

\bibitem{unpublished11}
S.-N. Hong, I.~Mari\'c, D.~Hui, and G.~Caire, ``On the achievable rates of
  multihop virtual full-duplex relay channels,'' 2015, arXiv:1501.06440.

\bibitem{inproceedings11}
S.~H. Lim, K.~T. Kim, and Y.-H. Kim, ``Distributed decode-forward for
  multicast,'' in \emph{Proc. Int. Symp. Inf. Theory (ISIT)}, 2014, pp.
  636--640.

\bibitem{unpublished9}
S.-H. Lee and S.-Y. Chung, ``Noisy network coding with partial {DF},'' 2015,
  arXiv:1505.05435.

\bibitem{article39:article40}
A.~Sendonaris, E.~Erkip, and B.~Aazhang, ``User cooperation diversity - part
  {I} \& {II}.'' \emph{IEEE Trans. Commun.}, vol.~51, no.~11, pp. 1927--1948,
  Nov. 2003.

\bibitem{article36}
S.~Ma, Y.~Yang, and H.~Sharif, ``Distributed {MIMO} technologies in cooperative
  wireless networks,'' \emph{IEEE Commun. Mag.}, vol.~49, no.~5, pp. 78--82,
  May 2011.

\bibitem{article42}
Y.~Jing and H.~Jafarkhani, ``Network beamforming using relays with perfect
  channel information,'' \emph{IEEE Trans. Inf. Theory}, vol.~55, no.~6, pp.
  2499--2517, Jun. 2009.

\bibitem{book3}
A.~El~Gamal and Y.-H. Kim, \emph{Network Information Theory}.\hskip 1em plus
  0.5em minus 0.4em\relax Cambridge Univ. Press, 2012.

\bibitem{book1}
T.~Cover and J.~Thomas, \emph{Elements of Information Theory, Second
  Edition}.\hskip 1em plus 0.5em minus 0.4em\relax Wiley-Interscience, 2006.

\end{thebibliography}

\end{document}